\pdfoutput=1

\documentclass{article} %
\usepackage{paper_style,times}

\usepackage{amsmath,amsfonts,bm}

\def\eqref#1{equation~\ref{#1}}

\def\1{\bm{1}}

\DeclareMathAlphabet{\mathsfit}{\encodingdefault}{\sfdefault}{m}{sl}
\SetMathAlphabet{\mathsfit}{bold}{\encodingdefault}{\sfdefault}{bx}{n}

\usepackage{hyperref}
\usepackage{url}
\usepackage{booktabs}
\usepackage{array}
\usepackage{longtable}
\usepackage{multirow}
\usepackage{graphicx}
\usepackage{wrapfig}
\usepackage{afterpage}
\usepackage{subcaption}
\usepackage{listings}
\usepackage{tcolorbox}
\tcbuselibrary{breakable,skins}
\usepackage{etoc}

\title{GPUPhysBench: Benchmarking Coding Agents for Correct and Efficient GPU Physics Simulation}

\author{Yuchen Sun, Jinjin He, Sinan Wang, Bo Zhu \\
Georgia Institute of Technology}

\begin{document}
\etocdepthtag.toc{mtchapter}

\maketitle

\begin{abstract}
Writing fast GPU code for physical simulation is difficult: implementations must preserve numerical accuracy while handling irregular data access, synchronization, and iterative solvers. We introduce GPUPhysBench, a benchmark of 50 tasks testing whether coding agents can meet these demands. Tasks cover fluids, deformable solids, and granular materials, from individual simulation operators to complete simulators. Agents write, compile, test, and optimize GPU code with access to a NVIDIA GPU under fixed time budgets. We report pass rates and runtime performance relative to expert-optimized reference implementations. In a single-attempt evaluation of six frontier model-harness pairs, the two strongest pass all 50 tasks, but even the fastest reaches at least $0.9\times$ the reference speed on only 22\% of them, and no submission is more than 5\% faster than the reference. The largest gaps arise in collision detection, constraint solving, and iterative solvers. GPUPhysBench brings physical simulation workloads to coding-agent evaluation, testing both the ability to implement numerical methods correctly and the ability to make them run efficiently.
\end{abstract}

\section{Introduction}
Modern physical simulation and machine learning rely on GPUs for large-scale workloads, but realizing this performance requires carefully optimized CUDA implementations that demand expertise in numerical algorithms, parallel programming, and hardware-aware optimization. LLM-based coding agents~\citep{chen2026avo, wei2025astra} offer an opportunity to automate this labor-intensive process, but evaluating them requires benchmarks that measure both correctness and speed.

Physical simulation is an important domain. Beyond long-standing uses in video games~\citep{miles2014unified}, aerospace engineering, and manufacturing, it offers a scalable and controllable source of physically plausible interaction data for world models and embodied intelligence~\citep{viktor2021isaac}, where real-world data collection is costly and slow. This demand calls for simulators that are both accurate and efficient, increasing the value of automating their GPU implementation.

Existing benchmarks do not cover this setting. Simulation-oriented evaluations emphasize numerical correctness or rely on CPU-based scientific software and established solver libraries~\citep{somasekharan2025cfdllmbench, hang2026pdeagentbench}, while benchmarks for efficient GPU code are dominated by machine-learning operators~\citep{ouyang2025kernel, guan2026tritongym}; even CUDABench~\citep{zhu2026cudabench} includes only a handful of elementary numerical examples. Unlike regular tensor operators, simulators combine structured grids with particles, meshes, and dynamically evolving neighborhoods, so their bottlenecks include irregular memory access, atomic contention~\citep{gao2018gpumpm}, neighbor search, and load imbalance. Optimizing them also requires numerical reasoning: agents must preserve the prescribed discretization and stability properties, and in solver-dominated tasks the choice of algorithm and preconditioner determines runtime through convergence.

Existing benchmarks are also limited in how they evaluate generated code. Most rely on single-turn generation or short, fixed loops of generation, verification, and profiling. These assess local kernel synthesis but not the ability of modern coding agents to plan, modify files, compile and run programs, diagnose failures, and improve an implementation over an extended trajectory. Evaluating such agents requires a controlled environment that preserves their autonomy while standardizing resource budgets and hardware access.

To address both gaps, we introduce GPUPhysBench, a benchmark of 50 coding tasks spanning classical methods for fluid dynamics, deformable solids, and granular materials. Every task includes standardized test data and a reference implementation that human experts have numerically validated and optimized. We evaluate frontier models in full-featured coding-agent harnesses, such as Codex CLI and Claude Code, inside a sandbox that isolates references and evaluators and fixes time budgets and hardware. This design measures pass rates and runtime relative to expert implementations while capturing the long-horizon, tool-using optimization of complete coding-agent systems.

\section{Related Work}

\paragraph{GPU Kernel Generation} A line of benchmarks studies how well LLMs can write high-performance GPU kernels for ML workloads. KernelBench \citep{ouyang2025kernel} and MultiKernelBench \citep{wen2025multikernelbench} score generated kernels on correctness and speedup over PyTorch baselines. TritonBench \citep{li2025tritonbench}, Geak \citep{wang2025geak}, and TritonGym \citep{guan2026tritongym} target the Triton DSL \citep{tillet2019triton}. Beyond one-shot generation, agentic systems iteratively optimize kernels using compilation, testing, and profiling feedback. They explore the optimization space through multi-agent refinement loops \citep{wei2025astra, sun2026kernelskill, zhang2026accelopt}, tree search \citep{dong2025stark, cao2026ksearch}, or evolutionary search with LLM-based variation operators \citep{chen2026avo, liao2025kernelevolve, yoo2026mkevolve}.

\paragraph{LLMs for Scientific Computing} Several benchmarks test LLM-written scientific code: SciCode \citep{tian2024scicode} curates research coding problems across the natural sciences, while CFDLLMBench \citep{somasekharan2025cfdllmbench} and PDEAgent-Bench \citep{hang2026pdeagentbench} check generated CFD and PDE solvers for accuracy and efficiency. Beyond benchmarks, agentic systems apply LLMs to scientific computing workflows: they generate PDE solver code and refine it with execution feedback \citep{li2025codepde, dong2026autopde}, automate end-to-end OpenFOAM workflows \citep{yue2025foamagent, yue2025foamagent20}, or tackle neighboring tasks such as PDE control \citep{soroco2025pdecontroller}, discovery \citep{luo2025llm4pd}, and reduced-order modeling \citep{wang2026opinfllm}.

\section{Benchmark}
\subsection{Task Design and Coverage}
\textsc{GPUPhysBench} comprises 50 coding tasks in fluid dynamics, deformable solids, and granular materials, drawn from well-established methods in physical simulation. We select tasks that represent widely used algorithms, pose nontrivial numerical and GPU optimization challenges, and support automatic correctness verification and reliable performance measurement. Using established methods provides clear mathematical specifications and interpretable failure modes while leaving agents substantial freedom in algorithmic and low-level implementation. The fluid tasks use Eulerian grid solvers~\citep{zehnder2018reflection}, particle-in-cell/fluid-implicit-particle (PIC/FLIP) and affine particle-in-cell (APIC) methods~\citep{jiang2015apic}, position-based fluids (PBF)~\citep{miles2013pbf}, and the lattice Boltzmann method (LBM)~\citep{li2026lbm}. The solid and granular tasks use mass-spring systems~\citep{liu2013ms}, the finite element method (FEM)~\citep{sifakis2012fem}, the material point method (MPM)~\citep{Stomakhin2013snow}, extended position-based dynamics (XPBD)~\citep{Miles2016xpbd}, continuous collision detection (CCD)~\citep{Brochu2022ccd}, and the discrete element method (DEM)~\citep{lu2022dem}. Appendix~\ref{app:task_list} describes the covered methods and lists all tasks with their categories.

The tasks operate on structured grids, particles, lattices, and meshes, covering GPU computation patterns such as stencils, iterative linear solves, atomic scatter and gather, irregular neighborhood interactions, constraint projection, and collision processing. Operator tasks isolate performance-critical operations for fine-grained analysis, while full-simulator tasks require agents to integrate multiple stages into a complete, efficient simulator that preserves numerical behavior.

\begin{figure}[t]
\centering
\captionsetup[subfigure]{font=small,justification=centering}
\begin{subfigure}[t]{0.32\textwidth}
\centering
\includegraphics[width=\linewidth]{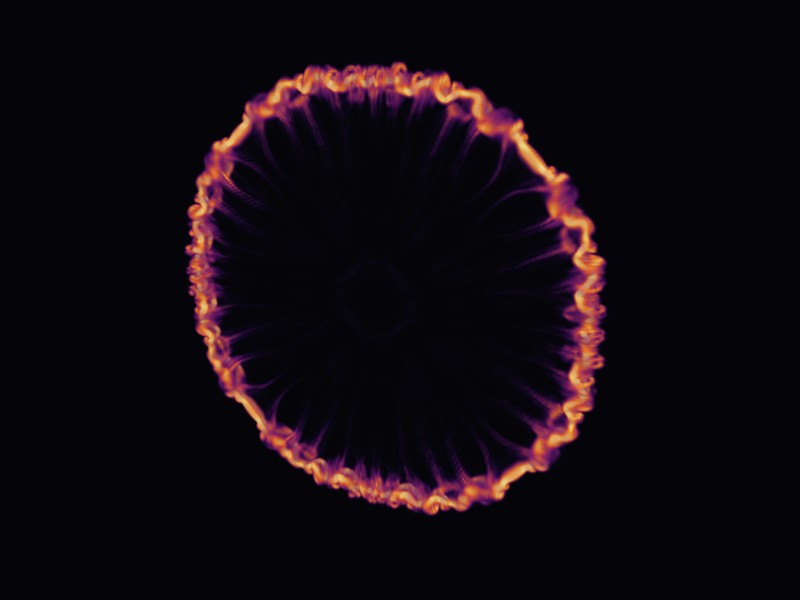}
\caption{Vortex-ring collision (\nolinkurl{mc_r})}
\end{subfigure}\hfill
\begin{subfigure}[t]{0.32\textwidth}
\centering
\includegraphics[width=\linewidth]{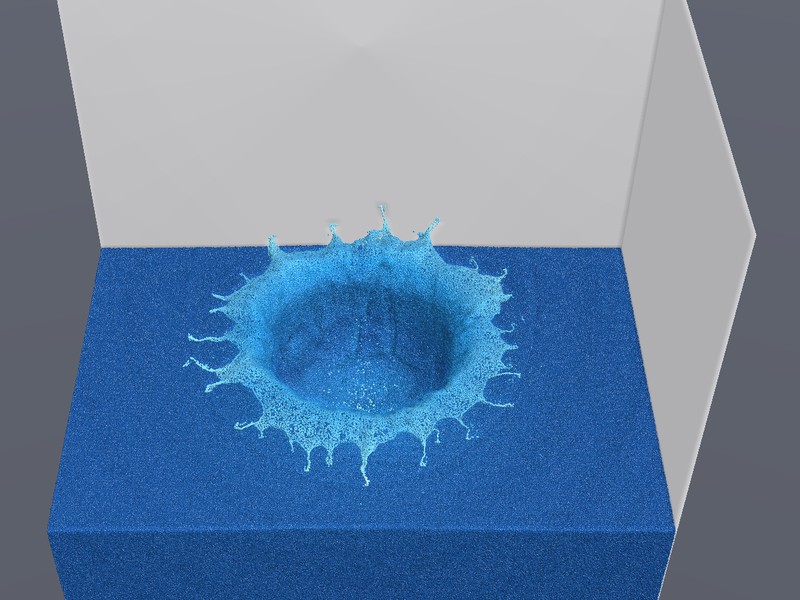}
\caption{Water drop (\nolinkurl{pic_flip})}
\end{subfigure}\hfill
\begin{subfigure}[t]{0.32\textwidth}
\centering
\includegraphics[width=\linewidth]{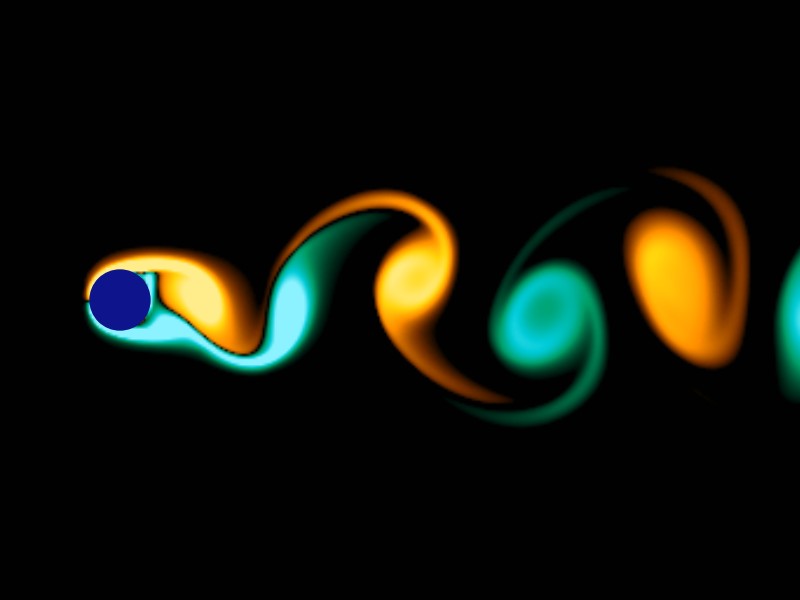}
\caption{K\'arm\'an vortex street (\nolinkurl{stable_fluids})}
\end{subfigure}

\medskip
\begin{subfigure}[t]{0.32\textwidth}
\centering
\includegraphics[width=\linewidth]{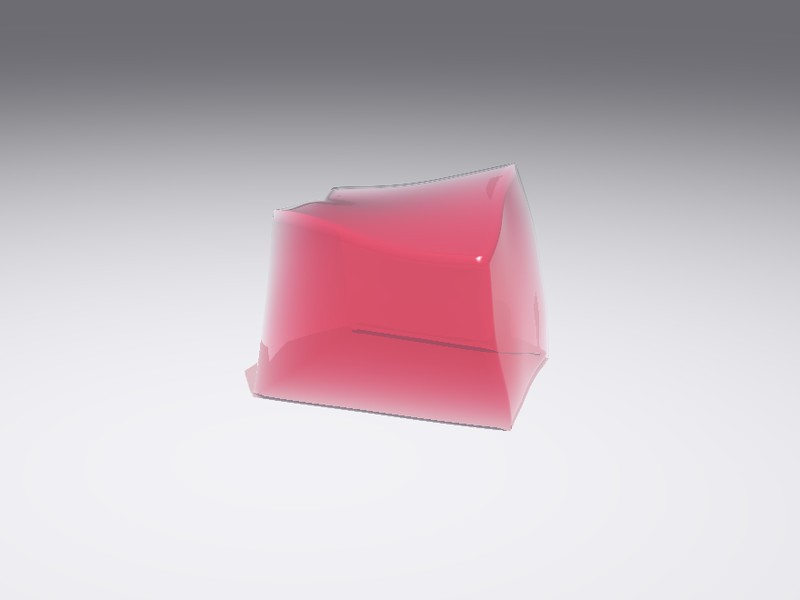}
\caption{Jelly cube (\nolinkurl{fem_explicit})}
\end{subfigure}\hfill
\begin{subfigure}[t]{0.32\textwidth}
\centering
\includegraphics[width=\linewidth]{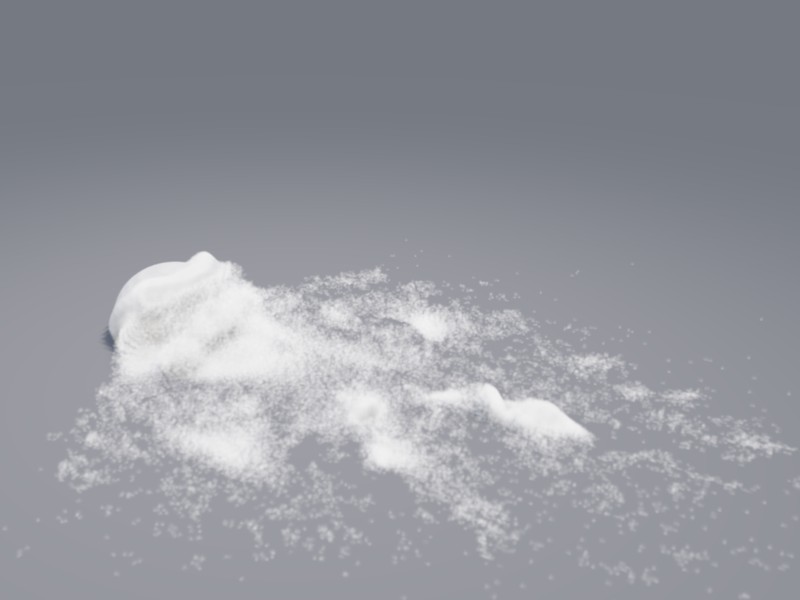}
\caption{Snowball (\nolinkurl{mpm_explicit})}
\end{subfigure}\hfill
\begin{subfigure}[t]{0.32\textwidth}
\centering
\includegraphics[width=\linewidth]{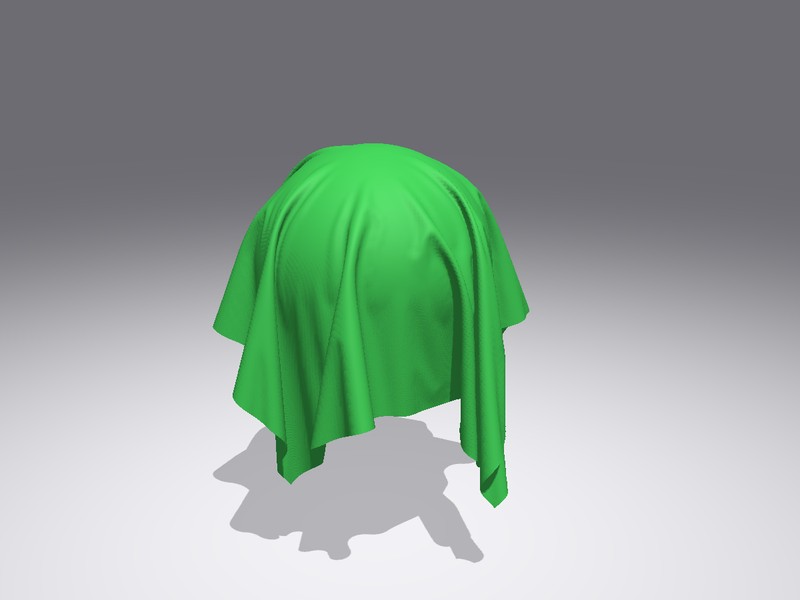}
\caption{Cloth on a sphere (\nolinkurl{xpbd})}
\end{subfigure}
\caption{Multi-step simulations driven by GPUPhysBench reference implementations.}
\label{fig:simulations}
\end{figure}

Each task comes with an expert-optimized CUDA reference implementation that defines both the correctness baseline and the performance target. We build the references from open-source, high-performance simulation code released with SIGGRAPH papers and courses, such as Fast UAAMG~\citep{shao2022uaamg}, GPUMPM~\citep{gao2018gpumpm}, and an LBM course~\citep{li2026lbm}; this code is written in C++, earlier versions of CUDA, and NVIDIA Warp~\citep{warp2022}. Claude Fable translates it into CUDA, and human experts then rewrite and optimize the result for modern GPUs. Although the upstream code may appear in the pretraining data of the evaluated models, the expert-optimized references are never visible to agents, and the performance gaps in our experiments (Section~\ref{sec:main_results}) indicate that recalling the upstream code does not by itself reach reference performance. We validate each reference numerically against a serial baseline and run complete multi-step simulations to confirm that it produces physically valid behavior (Figure~\ref{fig:simulations}; Appendix~\ref{app:simulation_visualization}). The references are also competitive with established GPU libraries: on the tasks' public inputs, they are $9.0\times$ faster than AMGX~\citep{naumov2015amgx} on the Poisson solve and $2.6$--$7.0\times$ faster than ports of Warp and Taichi~\citep{hu2019taichi} examples (Appendix~\ref{app:library_comparison}).

\subsection{Task specification}

In each task, an agent implements a specified simulation computation in CUDA and minimizes its GPU execution time subject to numerical correctness requirements (Figure~\ref{fig:gpuphysbench_overview}). It receives a natural-language prompt and a workspace with starter code and a Python driver, while the references and correctness evaluators are withheld.

\afterpage{\begin{figure}[t]
    \centering
    \includegraphics[width=\textwidth]{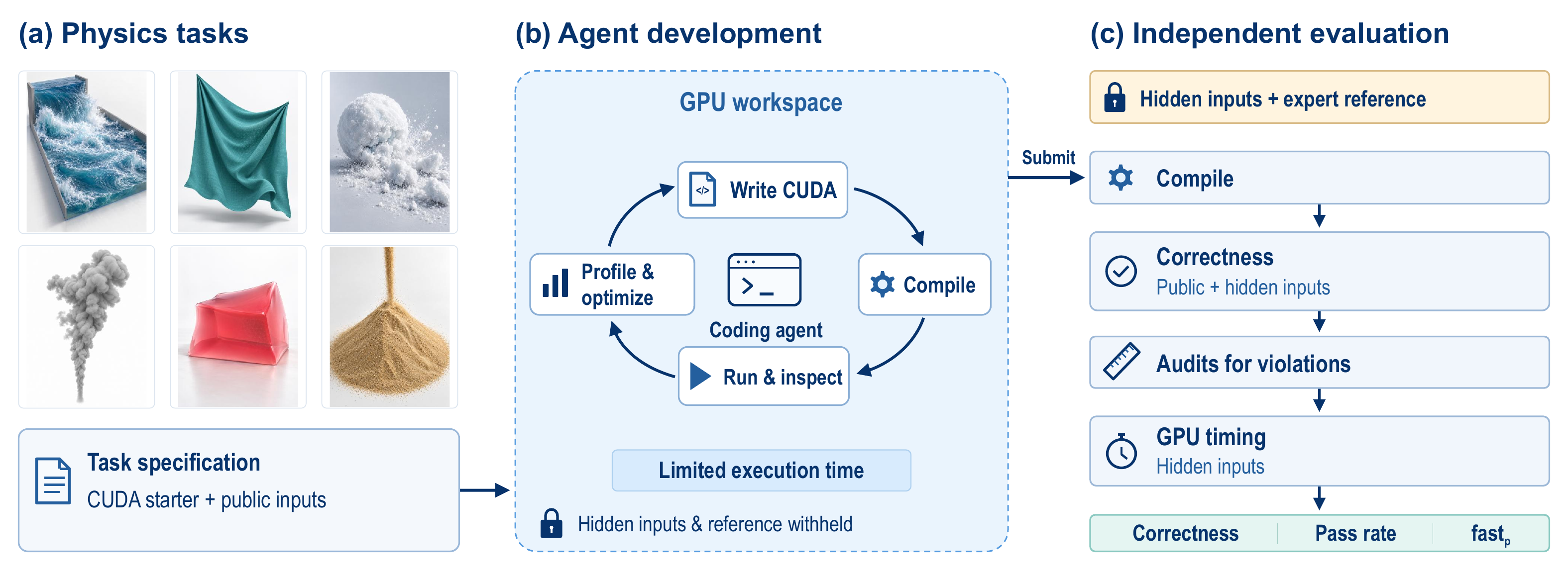}
    \caption{Overview of GPUPhysBench. Agents implement and optimize CUDA physics code under a time budget, without access to hidden inputs, references, or evaluators. Submissions are rebuilt, checked on public and hidden inputs, audited, and timed against expert references.}
    \label{fig:gpuphysbench_overview}
\end{figure}
}

\paragraph{Coding-agent harness.}
We evaluate agents in production-grade coding harnesses, such as Claude Code, that support file editing and tool use. Prior evaluations may sample or refine over multiple API calls, but each code-generating call emits a complete implementation, both in KernelBench~\citep{ouyang2025kernel} and in the scripted multi-call workflows evaluated by TritonGym~\citep{guan2026tritongym}. In our tests, generating a complete implementation in one call breaks down for large simulators, such as semi-implicit MPM, because the model's reasoning and code together can exceed the per-call token limit. Persistent harnesses instead let agents build a simulator incrementally, interleaving edits with compilation, testing, and optimization in their own order within the time budget.

\paragraph{Task input and environment.}
The prompt specifies the numerical method, including the governing equations or update rules, discretization, boundary conditions, and convergence criteria where applicable, together with the input and output arrays, scalar parameters, performance objective, and execution constraints. The workspace contains a CUDA source file, an xmake build file, and a Python driver, and the environment provides a GPU and the tooling to build and run the extension. The driver constructs the public inputs and runs the compiled module, so the agent can inspect inputs, outputs, and timing, but it provides no reference outputs or correctness verdicts. Online access is prohibited.

\paragraph{Starter code and required output.}
The starter code defines a task-specific class whose pybind11 bindings and \texttt{input()}/\texttt{exec()}/\texttt{output()} methods are fixed: they upload the inputs to device buffers, time the computation, and copy the output back. The agent implements \texttt{allocate()}, \texttt{compute()}, and \texttt{release()}, and may add CUDA kernels, internal buffers, and its own data layout; any layout conversion or preprocessing runs inside the timed \texttt{compute()}, while the untimed \texttt{allocate()} may only allocate memory from shapes and scalar parameters. The fixed code and module interface must be preserved, and build changes are limited to compilation flags. Appendix~\ref{app:example_task} gives the complete prompt and starter code for a Neumann Poisson task.

\paragraph{Numerical requirements.}
All floating-point computation and storage must use FP32. A submission must pass task-specific checks, such as relative $\ell_2$ error or solver convergence, on the outputs of its timed calls for both public and hidden inputs. The hidden inputs are withheld during development. They keep the problem size but change the data through different random seeds and initial conditions and, for some tasks, different geometry or solver coefficients (Appendix~\ref{app:task_inputs}). Any computational strategy is allowed as long as it meets these accuracy requirements.

\paragraph{Correctness tolerances.}
Most checks compare the submission's output with the reference output by relative $\ell_2$ error on the quantity the step changes; for example, particle positions are compared by their displacement over the step, so that large absolute coordinates cannot hide an error in the update. Linear and Newton solves instead require the true residual of the returned solution to fall below the prescribed solver tolerance. Each tolerance is set above the variation between correct FP32 implementations, which differ in rounding, reduction order, and the order of atomic accumulation, and well below the effect of plausible implementation errors, such as a missing term or a flipped sign. Appendices~\ref{app:error_margins} and~\ref{app:tolerance_validation} confirm both bounds: correct outputs stay well below their tolerances and erroneous ones far above them, so moderately different thresholds would not change any outcome.

\paragraph{Performance objective.}
The objective is to minimize GPU execution time while satisfying the numerical specification. CUDA events measure all GPU work performed during \texttt{exec()}, including auxiliary computation and solver iterations, while input upload and output download are excluded through the separate interface methods. The generated and reference implementations are evaluated on identical inputs under the same timing protocol.

\paragraph{Time budget.}
The agent receives a wall-clock budget of 30 minutes for operator tasks and 60 minutes for full-simulator tasks, covering implementation, compilation, testing, and optimization. The prompt states the budget and absolute deadline. The agent may independently check the current time using the \texttt{date} command and compare it with the deadline to decide whether to continue optimizing or finalize its submission. At most one agent, including the main agent and any subagents, may be active at a time. Any delegated execution shares the same task-level wall-clock budget. The run is terminated when the budget expires, and the submitted source and build configuration present when the agent finishes or reaches the deadline are used for evaluation.

\subsection{Anti-cheating}
\paragraph{File-system isolation.}
Each task runs in a Docker container, in a separate working directory that contains only the agent-facing files. The agent runs as an unprivileged user whose permissions block access to the benchmark project, including references and evaluators, and the harness verifies this isolation before launch. For scoring, the harness restores the fixed starter-code sections and rebuilds the submission in a clean directory with the protected driver and evaluator, ignoring any prebuilt module left by the agent.

\paragraph{Execution rules and post-run auditing.}
Agents may not access online resources, run agents in parallel, or move computation outside the timed region; built-in web search is disabled where the harness supports it. Static checks enforce the interface, build, and allocation restrictions, and Claude-Opus-5 audits each run's logs and code for network access, parallel agents, and untimed computation, using the same prompts for every system (Appendix~\ref{app:auditor}). Any violation counts as a failure, regardless of correctness or speed.

\subsection{Metrics}
\label{sec:metrics}

We evaluate agents using correctness, pass rate, and performance relative to the reference. All metrics are computed over all $N$ benchmark tasks using the final submission from each task attempt.

\paragraph{Correctness.}
Correctness is the fraction of tasks whose submission builds, runs, and passes all numerical checks on both public and hidden inputs, regardless of audit outcomes.

\paragraph{Pass Rate.}
Pass rate is the fraction of tasks whose submission is correct and also passes all anti-cheating audits; we write $v_i=1$ for such a passing submission and $v_i=0$ otherwise.

\paragraph{Performance.}
Following KernelBench~\citep{ouyang2025kernel}, we use $\mathrm{fast}_p$ to measure the fraction of tasks that pass and achieve a speedup greater than a threshold $p$. For a passing submission ($v_i=1$), its speedup is
\begin{equation}
    s_i = \frac{t_i^{\mathrm{ref}}}{t_i^{\mathrm{gen}}},
\end{equation}
where $t_i^{\mathrm{ref}}$ and $t_i^{\mathrm{gen}}$ are the GPU execution times of the reference and generated implementations, measured on the hidden inputs using the same hardware and timing protocol. We set $s_i=0$ for failed submissions. For $p \geq 0$,
\begin{equation}
    \mathrm{fast}_p = \frac{1}{N}\sum_{i=1}^{N} v_i\,\mathbf{1}[s_i > p].
\end{equation}
We report $p\in\{0.5,0.9,1.05\}$. Failed submissions remain in the denominator. $\mathrm{fast}_{1.05}$ counts passing implementations that outperform the reference by more than 5\%. We use this threshold instead of $p=1$ because a speedup just above $1$ is within timing noise. Each unchanged reference is timed 13--14 times across our evaluation sessions, and these timings vary with a median coefficient of variation of 0.9\% per task (at most 3.0\%) and a median max-to-min ratio of $1.03$.

\section{Experiments}

We evaluate coding agents on GPUPhysBench to assess their ability to produce numerically correct and efficient GPU simulation programs. We compare six model-harness pairs against expert implementations in terms of correctness, pass rate, and runtime performance, and analyze the generated programs to identify the factors that contribute to the remaining performance gap.

\subsection{Setup}
\label{sec:setup}
All agent development and evaluation run on a single NVIDIA GeForce RTX 4090 (Ada Lovelace), so agents tune on the same GPU that scores them, inside an Ubuntu-based CUDA Docker image with Python, pybind11, and xmake. For each implementation and input set, we report the minimum of five timed calls, each in a fresh process after three warm-up calls on the other input set; speedups use the hidden-input timings.

We evaluate six frontier models, using their corresponding coding harnesses where possible. We pair Claude-Opus-5 with Claude Code, GPT-5.6-Sol with Codex CLI, Gemini-3.5-Flash with Gemini CLI, DeepSeek-V4.1-Flash with DeepSeek Harness, and Qwen-3.8-Max with Qwen Code. For GLM-5.3, we use OpenCode instead of ZCode, since the latter is a desktop development environment rather than a CLI harness. We refer to the six systems by their model family: Opus, GPT, Gemini, DeepSeek, Qwen, and GLM.
We configure reasoning effort to high for all applicable harnesses except Gemini CLI, which does not expose a corresponding effort-level setting. Each system receives one attempt per task in Table~\ref{tab:main_results}; Appendix~\ref{app:agent_cost} reports development time and token usage.

\subsection{Main Result}
\label{sec:main_results}
\begin{table*}[t]
\centering
\caption{Results on GPUPhysBench over 50 tasks, with one attempt per task. The best result in each column is underlined.}
\label{tab:main_results}

\small
\setlength{\tabcolsep}{3pt}
\renewcommand{\arraystretch}{1.08}

\begin{tabular*}{\textwidth}{@{\extracolsep{\fill}}llccccc@{}}
\toprule
\textbf{Model} &
\textbf{Harness} &
\textbf{Correctness} $\uparrow$ &
\textbf{Pass Rate} $\uparrow$ &
$\mathbf{fast}_{0.5}$ $\uparrow$ &
$\mathbf{fast}_{0.9}$ $\uparrow$ &
$\mathbf{fast}_{1.05}$ $\uparrow$ \\
\midrule

Claude-Opus-5 & Claude Code & \underline{100\%} & \underline{100\%} & \underline{66\%} & \underline{22\%} & 0\% \\
GPT-5.6-Sol & Codex CLI & \underline{100\%} & \underline{100\%} & 42\% & 16\% & 0\% \\
Gemini-3.5-Flash & Gemini CLI & 88\% & 88\% & 28\% & 16\% & 0\% \\
DeepSeek-V4.1-Flash & DeepSeek Harness & 86\% & 86\% & 38\% & 16\% & 0\% \\
Qwen-3.8-Max & Qwen Code & 52\% & 52\% & 34\% & 14\% & 0\% \\
GLM-5.3 & OpenCode & 88\% & 86\% & 34\% & 16\% & 0\% \\

\bottomrule
\end{tabular*}
\end{table*}

Table~\ref{tab:main_results} summarizes the performance of six model-harness pairs across all 50 GPUPhysBench tasks. These results assess complete coding systems: each agent must translate a numerical specification into CUDA code, resolve implementation issues, and optimize execution within a fixed time budget. We report numerical correctness separately from audit-compliant success and execution efficiency.

\paragraph{Observation 1: Frontier models can correctly implement fully specified numerical methods.}
Opus and GPT both pass all numerical checks and audits on all 50 tasks, including full simulators that require multiple numerical stages to work together, achieving 100\% correctness and pass rate. Gemini reaches 88\% on both metrics, DeepSeek 86\%, and Qwen 52\%, while GLM reaches 88\% correctness and an 86\% pass rate. This result should be read in light of the task format: each prompt prescribes the numerical method, discretization, boundary conditions, and convergence criteria, so the tasks test faithful implementation of a given method rather than the choice of physical model. For the strongest systems, correctness is therefore close to saturation, and performance is the axis on which GPUPhysBench separates them.

\paragraph{Observation 2: Simulation performance still lags behind expert references.}
Despite matching Opus in correctness and pass rate, GPT achieves $\mathrm{fast}_{0.5}=42\%$, compared with Opus's 66\%, revealing a substantial gap in execution efficiency; Opus is faster than GPT on 38 of the 50 tasks. Even for Opus, 17 of the 50 passing submissions take at least twice as long as the reference. Opus leads at $\mathrm{fast}_{0.9}=22\%$, while the other systems reach 14\%--16\%; their $\mathrm{fast}_{0.9}$ successes come only from regular, memory-bound local grid and lattice computations, on which nearly all systems come close to the reference, whereas Opus also approaches the reference on a few particle and mesh tasks with irregular access. No passing submission is more than 5\% faster than the reference ($\mathrm{fast}_{1.05}=0$ for every system). Thus, reliable numerical implementation does not yet translate into performance comparable to expert references on most tasks.

\newpage
\subsection{Performance by Task Category}
\label{sec:category_results}
\begin{wrapfigure}{r}{0.55\textwidth}
\centering
\includegraphics[width=\linewidth]{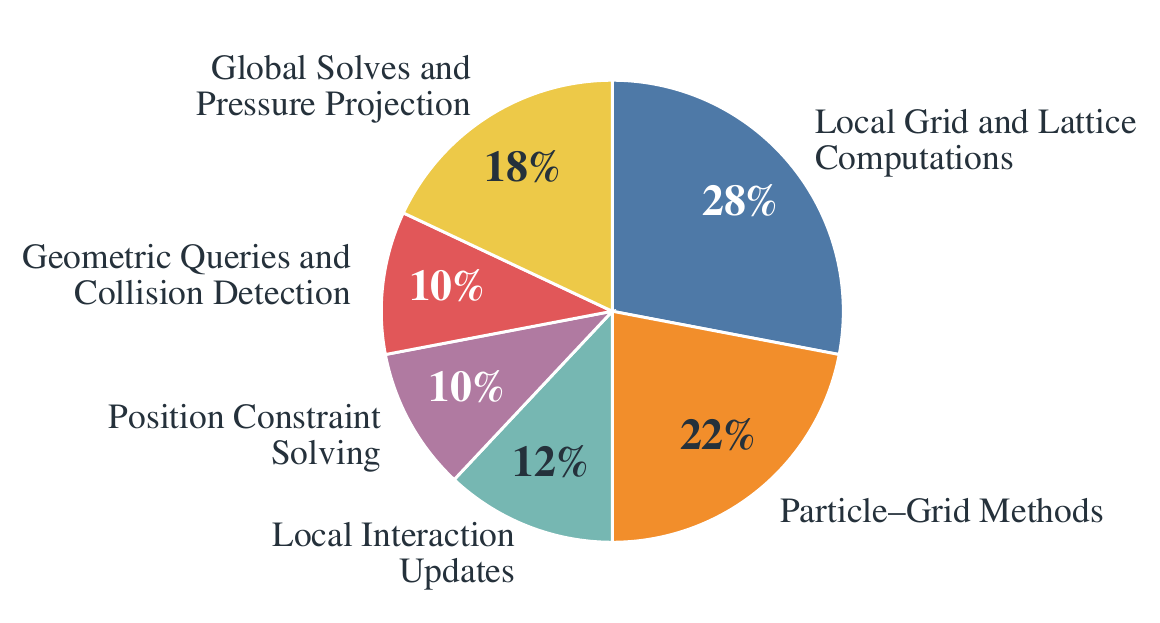}
\caption{Distribution of the 50 tasks across six computational categories.}
\label{fig:task_categories}
\end{wrapfigure}
To compare benchmark outcomes across computational structures, we group tasks into six categories according to their numerical methods and computational structure (Figure~\ref{fig:task_categories}). \emph{Local Grid and Lattice Computations} covers advection, LBM operations, and local surface-tension and phase-field updates. \emph{Particle-Grid Methods} includes transfer operators and complete PIC/FLIP, APIC, and MPM steps. \emph{Local Interaction Updates} covers explicit elasticity, contact-force evaluation, and local viscosity and vorticity velocity updates. \emph{Position Constraint Solving} contains PBF and XPBD constraint projections and complete steps. \emph{Geometric Queries and Collision Detection} comprises CCD and particle-based distance-field construction. \emph{Global Solves and Pressure Projection} includes Poisson and implicit viscosity solves, Newton and projective-dynamics elasticity, and grid-based fluid steps with pressure projection.

\afterpage{%
\begin{figure}[t]
\centering
\captionsetup[subfigure]{font=small,justification=centering}
\begin{subfigure}[t]{0.32\textwidth}
\centering
\includegraphics[width=\linewidth]{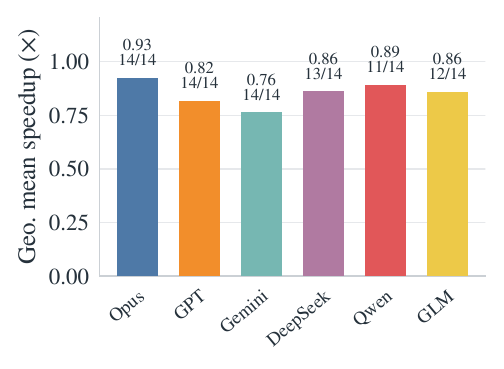}
\caption{Local Grid and Lattice Computations}
\label{fig:speedup_grid}
\end{subfigure}\hfill
\begin{subfigure}[t]{0.32\textwidth}
\centering
\includegraphics[width=\linewidth]{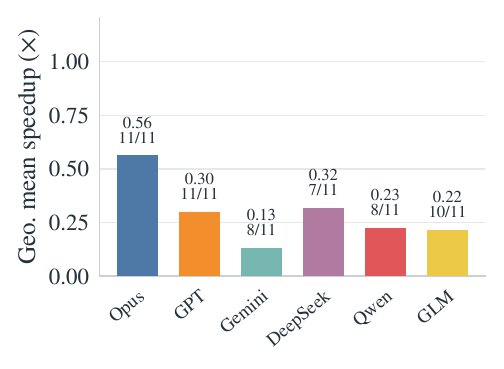}
\caption{Particle-Grid Methods}
\label{fig:speedup_particle_grid}
\end{subfigure}\hfill
\begin{subfigure}[t]{0.32\textwidth}
\centering
\includegraphics[width=\linewidth]{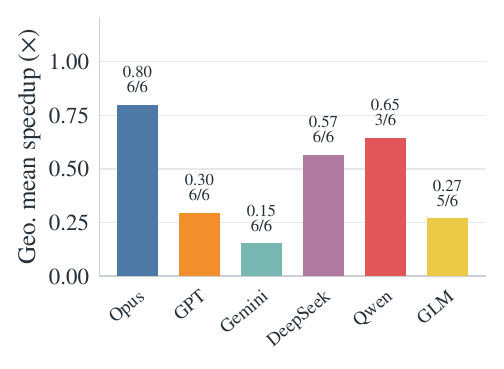}
\caption{Local Interaction Updates}
\label{fig:speedup_interaction}
\end{subfigure}

\medskip
\begin{subfigure}[t]{0.32\textwidth}
\centering
\includegraphics[width=\linewidth]{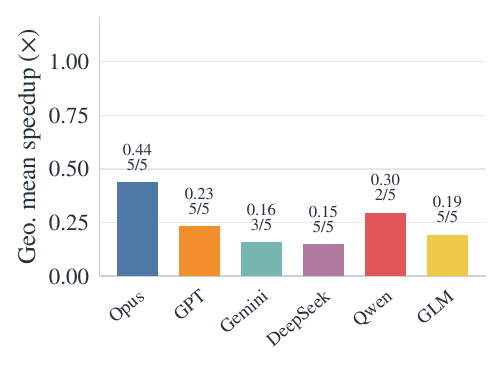}
\caption{Position Constraint Solving}
\label{fig:speedup_constraint}
\end{subfigure}\hfill
\begin{subfigure}[t]{0.32\textwidth}
\centering
\includegraphics[width=\linewidth]{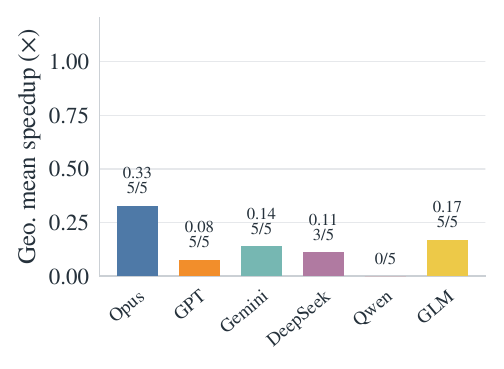}
\caption{Geometric Queries and Collision Detection}
\label{fig:speedup_geometry}
\end{subfigure}\hfill
\begin{subfigure}[t]{0.32\textwidth}
\centering
\includegraphics[width=\linewidth]{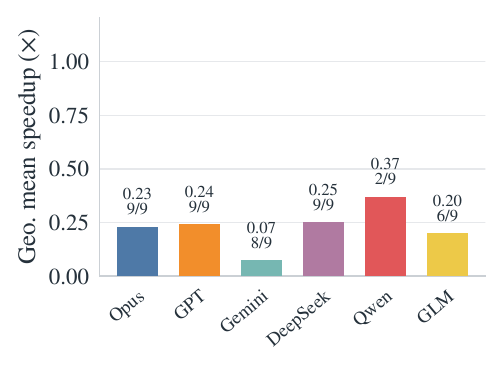}
\caption{Global Solves and Pressure Projection}
\label{fig:speedup_global}
\end{subfigure}
\caption{Geometric mean speedup relative to the expert reference on hidden inputs, over each system's passing tasks in each category. Labels give the mean and the number of passing tasks; a system with no passing task has no bar. Axis scales are shared across panels.}
\label{fig:category_speedups}
\end{figure}
}

Figure~\ref{fig:category_speedups} reports, for each system and category, the geometric mean speedup over the tasks the system passes, so speed is reported separately from success: Qwen's high mean on global solves covers only 2 of 9 tasks. Every system is fastest on local grid and lattice computations, and Opus leads in five of the six categories; on global solves, Opus, GPT, and DeepSeek lie between $0.23\times$ and $0.25\times$. Geometric queries and collision detection are the slowest on average, followed by global solves and position constraint solving, while local interaction updates vary the most across systems. Weighting each task family equally (Appendix~\ref{app:family_balanced}) leaves the ordering by $\mathrm{fast}_{0.5}$ unchanged but lowers the $\mathrm{fast}_{0.9}$ of every system other than Opus to 8\%, because their successes concentrate in the LBM family. Appendix~\ref{app:case_studies} examines two cases of generated code in detail.

\paragraph{Overall performance across categories.}
Agents perform best on local computations, particularly regular grid and lattice operations. All passing submissions with speedups above $1\times$ belong to the local grid and lattice category, and their gains are below 0.3\%, with a median speedup of $1.001\times$ and a maximum of $1.002\times$. These margins are smaller than the run-to-run variation of the timings (Section~\ref{sec:metrics}) and do not establish a performance advantage. Global solves and more complex workloads involving irregular data access, iterative updates, or coordination across multiple stages show larger performance gaps. Overall, agents are more successful at exploiting regular local parallelism than at jointly optimizing algorithmic choices, data organization, and execution across an entire simulation step.

\paragraph{Local Grid and Lattice Computations.}
Near-reference performance is concentrated in LBM, streaming, collision, and surface-tension tasks, where regular indexing and local arithmetic are readily mapped to GPU threads. Advection and Cahn--Hilliard updates show larger gaps and more variation across systems. In third-order advection, for example, Opus processes all three velocity components in one kernel, whereas GPT launches separate component kernels. These implementations highlight opportunities to reuse interpolation data and combine stages even within otherwise regular workloads.

\paragraph{Particle-Grid Methods.}
Spatial ordering alone does not ensure efficient transfers. In PIC/FLIP P2G, Gemini sorts particle indices but retains indirect particle reads and global atomics for each contribution. Opus packs particles into spatial tiles and accumulates in shared memory, reaching $0.556\times$ the reference speed versus Gemini's $0.134\times$. The reference further aggregates contributions by cell before merging them into shared memory (Appendix~\ref{app:case_p2g}). The remaining gap therefore involves both the cost of grouping particles and the granularity of accumulation; sorting or moving atomics into shared memory is only part of the optimization.

\paragraph{Local Interaction Updates.}
Performance varies substantially even among correct implementations of local interactions. Opus approaches reference performance on explicit FEM and matches it closely on DEM contact, while other systems often remain much slower. In DEM contact, both Opus and GPT pack particle records, but Opus traverses contiguous ranges in a spatial grid, whereas GPT searches hash buckets and filters candidates by cell coordinates. These differences highlight the importance of neighbor-search organization and candidate access beyond the arithmetic of the local force or velocity update.

\paragraph{Position Constraint Solving.}
PBF implementations already exploit spatial reordering: Opus rebuilds cell offsets and rearranges particles during each constraint iteration, reaching $0.815\times$ reference speed on the standalone PBF solve. Spring-based XPBD remains harder, with every system below $0.3\times$ on the standalone constraint solve. Opus accumulates spring corrections through atomics into packed particle buffers; GPT instead builds adjacency lists and gathers incident corrections. Both repeatedly exchange corrections and positions through global memory, leaving opportunities to reduce iteration traffic and improve reuse across constraints.

\paragraph{Geometric Queries and Collision Detection.}
All but one passing continuous-collision-detection submission remains below $0.5\times$ reference speed, with the best reaching $0.526\times$, despite using spatial acceleration. The implementations span different structures: Opus uses a spatial grid for vertex--vertex CCD, while GPT builds and traverses a bounding-volume hierarchy yet reaches only $0.007\times$; Gemini's vertex--face implementation sorts face--cell overlap pairs. These choices introduce different construction costs, candidate lists, and traversal patterns that must be optimized together. Particle distance-field construction generally performs better than CCD, but still trails the reference, suggesting that efficient candidate pruning and geometry access remain important across this category.

\paragraph{Global Solves and Pressure Projection.}
Category averages hide substantial differences in solver choice. Opus and Gemini both use Jacobi (diagonal) preconditioning for the Dirichlet Poisson solve, which Opus applies by symmetric rescaling; they reach only $0.049\times$ and $0.036\times$ reference speed, respectively. Opus fuses iteration kernels and keeps CG coefficients on the device, but these optimizations leave a large gap (Appendix~\ref{app:case_poisson}). GPT, GLM, and DeepSeek use multigrid preconditioners on the same task and reach $0.241\times$, $0.248\times$, and $0.303\times$, respectively. All three still trail the reference, underscoring the need to optimize convergence, memory traffic and synchronization.

\subsection{Ablation Study}
\label{sec:ablation}

\paragraph{Time budget.}
We evaluate Opus and GPT with one attempt at $0.5\times$, $1.0\times$, and $1.5\times$ the default time budget (Table~\ref{tab:ablation}). As the budget grows, Opus's $\mathrm{fast}_{0.5}$ rises steadily (54\%, 66\%, and 76\%), whereas GPT's changes little (36\%, 42\%, and 40\%). Correctness and pass rate remain at 100\% except for Opus at $0.5\times$, where both are 98\%. At $1.5\times$, two Opus submissions are more than 5\% faster than the reference, on DEM ($1.11\times$) and MPM grid-to-particle transfer ($1.07\times$). Thus, extra time primarily improves optimization, with larger gains for Opus in these runs. The default budget is rarely binding: at $1.0\times$, Opus and GPT use a median of 43\% and 50\% of it and never reach the deadline (Appendix~\ref{app:agent_cost}). The gains at $1.5\times$ therefore reflect how agents choose to use a longer stated deadline more than a lack of time.

\paragraph{Number of attempts.}
At the default budget, we run each model three times and compare the first run with the best of three, which keeps, for each task, the run with the highest hidden-input speedup among those that pass numerical checks and all audits. Correctness and pass rate reach 100\% for both models, although single runs pass 50, 49, and 50 tasks for Opus and 50, 47, and 49 for GPT. $\mathrm{fast}_{0.5}$ rises from 66\% to 76\% for Opus and from 42\% to 48\% for GPT, well beyond the run-to-run standard deviation of single-attempt $\mathrm{fast}_{0.5}$ (1.2 and 2.3 points; Appendix~\ref{app:repeated_runs}). $\mathrm{fast}_{0.9}$ rises from 22\% to 36\% for Opus, whereas GPT's increase from 16\% to 18\% is within run-to-run variation, since its second run alone reaches 18\%. Among runs that pass all numerical checks and audits, selecting by public-input speedup gives the same $\mathrm{fast}_{0.5}$ and $\mathrm{fast}_{0.9}$. This post-hoc comparison uses evaluator-only information to identify passing runs and compute reference-relative speedups. Only one run, from Opus on DEM ($1.13\times$), beats the reference by more than 5\% ($\mathrm{fast}_{1.05}$ of 2\% vs.\ 0\% for GPT). Additional attempts therefore improve peak performance rather than task coverage.

\begin{table}[t]
\centering
\caption{Ablations of time budget and number of attempts. Budgets are relative to the default of 30 minutes for operators and 60 minutes for full simulators. With three attempts, a task counts as solved if any run solves it, and $\mathrm{fast}_p$ uses the best passing speedup per task.}
\label{tab:ablation}
\small
\setlength{\tabcolsep}{2.5pt}
\renewcommand{\arraystretch}{1.08}
\begin{tabular*}{\textwidth}{@{\extracolsep{\fill}}cclccccc@{}}
\toprule
\textbf{Attempts} & \textbf{Time Budget} & \textbf{Coding Agent} & \textbf{Correctness} $\uparrow$ & \textbf{Pass Rate} $\uparrow$ & $\mathbf{fast}_{0.5}$ $\uparrow$ & $\mathbf{fast}_{0.9}$ $\uparrow$ & $\mathbf{fast}_{1.05}$ $\uparrow$ \\
\midrule
\multirow{6}{*}{1} & \multirow{2}{*}{$0.5\times$} & Opus & 98\% & 98\% & 54\% & 18\% & 0\% \\
 &  & GPT & 100\% & 100\% & 36\% & 16\% & 0\% \\
\cmidrule(lr){2-8}
 & \multirow{2}{*}{$1.0\times$} & Opus & 100\% & 100\% & 66\% & 22\% & 0\% \\
 &  & GPT & 100\% & 100\% & 42\% & 16\% & 0\% \\
\cmidrule(lr){2-8}
 & \multirow{2}{*}{$1.5\times$} & Opus & 100\% & 100\% & 76\% & 34\% & 4\% \\
 &  & GPT & 100\% & 100\% & 40\% & 18\% & 0\% \\
\midrule
\multirow{2}{*}{3} & \multirow{2}{*}{$1.0\times$} & Opus & 100\% & 100\% & 76\% & 36\% & 2\% \\
 &  & GPT & 100\% & 100\% & 48\% & 18\% & 0\% \\
\bottomrule
\end{tabular*}
\end{table}

\section{Conclusion}
We introduced GPUPhysBench, a benchmark of 50 GPU physics simulation tasks spanning individual operators and complete simulation steps. By evaluating coding agents in interactive development environments against numerical checks and expert reference implementations, GPUPhysBench measures both implementation correctness and execution efficiency. Our results show that frontier agents can correctly implement fully specified simulation methods, with the two strongest systems passing all tasks, while substantial performance gaps remain, particularly for collision detection, constraint solving, and iterative solvers. Analysis of generated code and development traces highlights the importance of numerical algorithm selection, data organization, and coordination across kernels. These findings motivate agents that reason jointly about numerical methods and GPU execution, and establish GPUPhysBench as a testbed for progress toward automated implementation of efficient physical simulators.

\section*{AI Use Statement}
We used generative AI tools to assist in constructing the benchmark, to draft sections of the paper, and to aid and polish the writing. In building the reference implementations, Claude Fable translated the upstream simulation code into CUDA, and human experts then rewrote and optimized the result. The authors carefully reviewed all AI-assisted work, including the benchmark tasks, the experimental results, and the text of the paper, and take full responsibility for the content of this work.

\section*{Reproducibility Statement}
We will release GPUPhysBench under an open-source license, including all 50 task prompts, starter code, drivers, reference implementations, correctness evaluators, the Docker environment, and the evaluation and audit scripts. Appendix~\ref{app:example_task} gives the complete prompt and starter code of one task, Appendix~\ref{app:task_inputs} the input sizes and correctness criteria of every task, and Appendix~\ref{app:auditor} the audit prompts.

\bibliography{references}
\bibliographystyle{paper_style}

\clearpage
\appendix
\etocdepthtag.toc{mtappendix}
\etocsettagdepth{mtchapter}{none}
\etocsettagdepth{mtappendix}{subsection}
\etocsettocstyle{\section*{Appendix Contents}}{}
\tableofcontents
\clearpage
\section{Example Task Prompt and Starter Code}
\label{app:example_task}

\newcommand{\examplemonofont}{\fontencoding{OT1}\fontfamily{cmtt}\selectfont}

We use the Neumann Poisson task to illustrate the agent's specification and CUDA interface. The complete prompt below reproduces the task specification, including the requirement that all algorithmic computation occur inside the timed \texttt{compute()}, and the constraints appended by the evaluation harness. Only run-specific deadlines are replaced by placeholders. Listing~\ref{lst:poisson_starter} reproduces the complete initial CUDA source, including its interface documentation, fixed interface methods, empty implementation hooks, timing scaffold, and pybind11 bindings. Appendix~\ref{app:poisson_driver} describes the Python driver that builds the task's inputs and runs a compiled module.

\subsection{Task Prompt}

\begin{tcolorbox}[
    enhanced,
    breakable,
    colback=black!3,
    colframe=black!60,
    colbacktitle=black!60,
    coltitle=white,
    boxrule=0.6pt,
    arc=1mm,
    left=8pt,right=8pt,top=6pt,bottom=6pt,
    fonttitle=\small\bfseries,
    fontupper=\fontsize{9}{10.5}\selectfont,
    title={Neumann Poisson Prompt},
    title after break={Neumann Poisson Prompt (continued)}
]
\setlength{\parindent}{0pt}
\setlength{\parskip}{4pt}
\renewcommand{\ttfamily}{\examplemonofont}
\par\texttt{\textless{}TASK\textgreater{}}\par
Implement a preconditioned Conjugate Gradient solver in CUDA, exposed to Python as a pybind11 extension module. Use xmake for compilation. Use FP32 (single-precision floating point) throughout the entire implementation. Do not use FP64 (double-precision floating point) for any computation or storage. Make the implementation as fast as you can: its GPU time is measured and reported, so optimize the CUDA code for performance. Do not tailor the implementation or its optimizations to the example input that run\_poisson\_neumann.py builds: it must be correct and efficient for any valid input as specified below, and must not rely on properties that happen to hold for that particular example.

\textbf{1. Linear System}\par
Solve the Poisson-type system A x = b on a uniform cell-centered grid of shape (res\_x, res\_y, res\_z); every input array (a\_diag, a\_x, a\_y, a\_z, b, is\_dof) and the output x have this shape. Only cells marked by the boolean array is\_dof are active degrees of freedom. Support arbitrary coefficients and active-cell configurations within the guarantees below, start from a zero initial guess, and iterate until the solution x you return -- after the zero-mean shift described below -- satisfies \textbar{}\textbar{}b - A x\textbar{}\textbar{}\_2 / \textbar{}\textbar{}b\textbar{}\textbar{}\_2 \textless{} tol, with A x taken over the active cells. This is the true residual of the returned x, not the residual carried along by the iteration's recurrences, which can drift from it in FP32.

The system has pure Neumann boundary conditions: the normal derivative is zero at both the grid boundary and interfaces with inactive cells. These conditions are already encoded in the input coefficients. Each active row sums to zero, up to FP32 rounding: its diagonal equals the sum of the magnitudes of its retained couplings. Removing a coupling also reduces the diagonal, so a missing neighbor means zero flux rather than a prescribed zero value.

A valid input also satisfies the following. The off-diagonal entries are zero or negative. A coupling to an inactive cell or to a cell beyond the grid is stored as zero, and every entry of an inactive cell -- a\_diag, a\_x, a\_y, a\_z and b -- is zero. Every active cell's diagonal is positive, though it may be very small. The tolerance tol is positive.

No boundary fixes the solution value, so A is singular. The active cells form one component, connected through non-zero couplings, giving a one-dimensional null space spanned by the vector that is one on active cells and zero elsewhere. The right-hand side sums to zero over active cells up to floating-point rounding; prevent the resulting null-space component from growing during iteration. Since the solution is determined only up to an additive constant, return the representative with zero mean over active cells, and set inactive cells to zero. A right-hand side of zero has the solution zero.

\textbf{2. Implementation and interface}\par
Fill in the implementation in the single file poisson\_neumann\_gen.cu and build it into the Python extension poisson\_neumann\_gen.so with the provided xmake.lua. The file already defines the \texttt{PoissonNeumann} class, its pybind11 bindings, and the module init; the comments above the class describe the input and output format -- what each array holds and the coefficient conventions -- and the comments inside it name each device buffer and member. The class already implements \texttt{input()}, \texttt{exec()} and \texttt{output()}: \texttt{input()} validates the arrays, stores the scalar parameters in members, copies each input array into a device buffer with the same layout, allocates a device buffer for the output, and then calls \texttt{allocate()}; \texttt{exec()} records a CUDA start event, calls \texttt{compute()}, synchronizes the device, records the stop event, and returns the elapsed GPU time in milliseconds; \texttt{output()} copies the output device buffer into the array it is given. Do not modify \texttt{input()}, \texttt{exec()}, \texttt{output()}, the destructor, the error helper, the bindings, or the members \texttt{input()} sets, and do not change what they do indirectly -- through macros, overloads or wrappers that redefine the CUDA calls or names they use, whether in the source file or through xmake.lua. The working directory also contains run\_poisson\_neumann.py, whose \texttt{run(folder, name)} builds the benchmark's input system and drives one compiled module, so you can call \texttt{run(".", "poisson\_neumann\_gen")} to exercise your own build and read back its timing and output. You may change the compilation flags in xmake.lua (optimization, architecture or register options, for example), but not its defines, include paths, forced includes, source files, targets or build scripts. For scoring, the extension is rebuilt from your source file and xmake.lua in a clean directory, with the fixed code above restored from the original skeleton -- so keep it, and the "===== Fixed" and "===== Implement below" marker comments, in place; a prebuilt .so you leave behind is not used. Each timed call runs in a fresh process, after warm-up calls on other inputs of the same kind, and the output of the timed call is the one checked. You may choose your own device-memory layout and internal data structures.

You implement the three private methods marked in the file, and may add members, device functions and kernels:

\begin{itemize}
\item \texttt{allocate()} allocates whatever extra memory \texttt{compute()} needs -- device memory with cudaMalloc or its variants, or pinned host memory with cudaMallocHost -- sized from the grid resolution and the scalar parameters. It may compute sizes and constants from those, and must do nothing else: no kernel launches, memory copies, memsets or other GPU work, no device, stream or kernel configuration (such as cudaFuncSetAttribute, cudaFuncSetCacheConfig, cudaDeviceSetLimit or an L2 access policy) and no creation of streams, events, CUDA graphs or library handles (do both in \texttt{compute()}), and nothing that reads the input data. Size queries that launch nothing (such as a CUB call with a null temporary buffer) and read-only queries of device or kernel properties are allowed. Storage whose size depends on the values in the input data, rather than only on the array shapes and scalar parameters, cannot be sized here: either reserve a bound computed from the shapes and scalar parameters, or allocate it inside \texttt{compute()}, where the allocation is timed.
\item \texttt{compute()} performs the entire solve. It reads the input device buffers, which it may overwrite, and writes the solution x into the output device buffer.
\item \texttt{release()} frees what \texttt{allocate()} allocated.
\end{itemize}

Do not read or write any files: nothing is loaded from disk, and the solution must be written into the output device buffer, from which \texttt{output()} copies it, rather than saved to x.npy or anywhere else.

Report CUDA errors as Python exceptions (e.g. throw std::runtime\_error, which pybind11 maps to a Python exception); the file's ck() helper does this for a CUDA call.

\textbf{3. Computation and timing}\par
Only the time \texttt{exec()} returns is scored, and it covers \texttt{compute()} alone, so everything you implement that belongs to the solve must run inside \texttt{compute()}, on every call -- including building the preconditioner, the zero-mean shift of the solution, any conversion of the input buffers into another layout, and any zero-initialization of buffers. Do not perform any part of it anywhere else: not in \texttt{allocate()} or \texttt{release()}, not in constructors, static or global initializers or the module init, not on a host thread that outlives \texttt{compute()}, and not by reusing results from an earlier call.

\par\texttt{\textless{}/TASK\textgreater{}}\par
\par\texttt{\textless{}CONSTRAINT\textgreater{}}\par
1. Do not perform any web searches or access any online resources.\par
2. Do not run agents in parallel. At most one agent may be working at any moment, counting yourself and any subagent you start: if you delegate, the subagent has to finish and hand back before you do anything else. Do not start two subagents in one step, do not start one while another is running, and do not put one in the background.\par
3. You have 30 minutes of wall-clock time, ending at \textless{}deadline\_iso\textgreater{} (Unix time \textless{}deadline\_epoch\textgreater{}). Run \texttt{date +\%s} and compare it against that number whenever you want to know how much is left. At the deadline the run is killed mid-action and whatever is on disk is what gets evaluated -- so get a working build in place early, and treat anything after that as optional improvement you can afford to lose.\par
\par\texttt{\textless{}/CONSTRAINT\textgreater{}}\par

\end{tcolorbox}

\clearpage
\subsection{Initial CUDA Code}
\label{app:poisson_starter}

The agent implements \texttt{allocate()}, \texttt{compute()}, and \texttt{release()} while preserving the fixed interface. The fixed \texttt{exec()} method times \texttt{compute()} and synchronizes the device before recording the stop event.

\begin{lstlisting}[
    language=C++,
    basicstyle=\fontsize{8}{9.3}\examplemonofont,
    keywordstyle=\color{blue!45!black},
    commentstyle=\color{black!55},
    stringstyle=\color{green!30!black},
    numbers=left,
    numberstyle=\tiny\color{black!45},
    numbersep=8pt,
    xleftmargin=18pt,
    frame=single,
    rulecolor=\color{black!35},
    backgroundcolor=\color{black!2},
    framesep=6pt,
    breaklines=true,
    breakatwhitespace=true,
    columns=fullflexible,
    keepspaces=true,
    showstringspaces=false,
    tabsize=4,
    captionpos=b,
    caption={Unmodified CUDA starter code for the Neumann Poisson task.},
    label={lst:poisson_starter}
]
#include <cuda_runtime.h>
#include <pybind11/numpy.h>
#include <pybind11/pybind11.h>

#include <stdexcept>
#include <string>

namespace py = pybind11;

namespace gen_impl
{

    // Solves a Poisson-type system A x = b on a uniform cell-centered grid of
    // shape (res_x, res_y, res_z), with a preconditioned Conjugate Gradient
    // iteration starting from a zero initial guess. Only a subset of the cells
    // are active degrees of freedom; the rest carry no equation.
    //
    // --- input format ------------------------------------------------------
    // Every array is 3D of shape (res_x, res_y, res_z), where [i, j, k] is the
    // cell at spatial index (x, y, z). The resolution is inferred from the
    // shapes; all six arrays must agree. a_diag, a_x, a_y, a_z and b are
    // float32; is_dof is bool.
    //
    //   a_diag[i,j,k]  the diagonal entry of A at cell (i, j, k)
    //   a_x[i,j,k]     the matrix entry coupling (i, j, k) to (i+1, j, k)
    //   a_y[i,j,k]     the matrix entry coupling (i, j, k) to (i, j+1, k)
    //   a_z[i,j,k]     the matrix entry coupling (i, j, k) to (i, j, k+1)
    //   b[i,j,k]       the right-hand side
    //   is_dof[i,j,k]  true when the cell is an active degree of freedom
    //
    // A is symmetric, so the entry coupling (i, j, k) to (i-1, j, k) is
    // a_x[i-1, j, k], and likewise along y and z. A coupling whose neighbour
    // would fall outside the grid is absent from the system: its entry, at the
    // last index along that axis, is present in the array but zero.
    //
    // Sign convention: a_x, a_y and a_z hold the actual signed matrix entries,
    // not positive coupling magnitudes. For the standard discrete Poisson
    // operator the off-diagonals are negative (e.g. -1 on a unit grid for an
    // interior face) and the diagonal is positive, so
    //
    //   (A x)[I] = a_diag[I] * x[I] + sum over neighbours of a_off * x[nbr]
    //
    // with a plus sign in front of the off-diagonal sum.
    //
    // Cells where is_dof is false hold no unknown: their row is not part of the
    // system, and they contribute nothing to an active cell's equation.
    //
    // --- pure Neumann boundary conditions -----------------------------------
    // The normal derivative is zero on every boundary, both at the edge of the
    // grid and at the interface with inactive cells. That makes every active
    // row sum to zero, up to FP32 rounding: a cell's diagonal is the sum of the
    // magnitudes of the couplings it keeps, so on a unit grid a cell with six
    // active neighbours has a_diag = 6 while a cell with three has a_diag = 3.
    // Where a coupling is dropped the diagonal drops with it -- a dropped
    // coupling means no flux across that face, not a known value beyond it.
    //
    // Since no boundary prescribes a value, A is singular: it annihilates any
    // function that is constant on the active set. The active cells form a
    // single connected component, so that null space is exactly
    // one-dimensional, spanned by the vector that is 1 on active cells and 0
    // elsewhere. Two things follow:
    //
    //   - b is compatible: it sums to zero over the active cells, so a solution
    //     exists. It is only compatible to within floating-point rounding, and
    //     the resulting component along the null space must not be allowed to
    //     grow as the iteration proceeds.
    //   - x is determined only up to an additive constant, which is why the
    //     output below is pinned to a particular representative.
    //
    // --- output format -----------------------------------------------------
    // One float32 array of the same shape (res_x, res_y, res_z) holding the
    // solution x, normalized to have zero mean over the active cells. The value
    // at every cell where is_dof is false must be set to zero. Nothing is
    // written to disk: the caller supplies the destination array and output()
    // fills it.

    // Throws a Python exception for a failed CUDA call.
    inline void ck(cudaError_t e, const char *what)
    {
        if (e != cudaSuccess)
            throw std::runtime_error(std::string(what) + ": " + cudaGetErrorString(e));
    }

    class PoissonNeumann
    {
    public:
        // ===== Fixed: do not modify input(), exec(), output(), the destructor,
        // ck(), the bindings or the members input() sets. =====

        // (1) input: validate the arrays, keep the scalars, copy each input
        // array into a device buffer of the same layout, allocate a device
        // buffer for each output, then call allocate().
        void input(py::array_t<float, py::array::c_style> a_diag,
                   py::array_t<float, py::array::c_style> a_x,
                   py::array_t<float, py::array::c_style> a_y,
                   py::array_t<float, py::array::c_style> a_z,
                   py::array_t<float, py::array::c_style> b,
                   py::array_t<bool, py::array::c_style> is_dof,
                   float tol)
        {
            auto bd = a_diag.request(), bx = a_x.request(), by = a_y.request(),
                 bz = a_z.request(), bb = b.request(), bm = is_dof.request();
            if (bd.ndim != 3)
                throw std::runtime_error("a_diag must have shape (res_x, res_y, res_z)");
            auto same = [&](const py::buffer_info &o) {
                return o.ndim == 3 && o.shape[0] == bd.shape[0] &&
                       o.shape[1] == bd.shape[1] && o.shape[2] == bd.shape[2];
            };
            if (!same(bx))
                throw std::runtime_error("a_x must have shape (res_x, res_y, res_z)");
            if (!same(by))
                throw std::runtime_error("a_y must have shape (res_x, res_y, res_z)");
            if (!same(bz))
                throw std::runtime_error("a_z must have shape (res_x, res_y, res_z)");
            if (!same(bb))
                throw std::runtime_error("b must have shape (res_x, res_y, res_z)");
            if (!same(bm))
                throw std::runtime_error("is_dof must have shape (res_x, res_y, res_z)");
            res_x_ = static_cast<int>(bd.shape[0]);
            res_y_ = static_cast<int>(bd.shape[1]);
            res_z_ = static_cast<int>(bd.shape[2]);
            num_cells_ = static_cast<size_t>(res_x_) * res_y_ * res_z_;
            tol_ = tol;

            const size_t fbytes = sizeof(float) * num_cells_;
            const size_t mbytes = sizeof(bool) * num_cells_;
            ck(cudaMalloc(&d_a_diag_, fbytes), "cudaMalloc a_diag");
            ck(cudaMalloc(&d_a_x_, fbytes), "cudaMalloc a_x");
            ck(cudaMalloc(&d_a_y_, fbytes), "cudaMalloc a_y");
            ck(cudaMalloc(&d_a_z_, fbytes), "cudaMalloc a_z");
            ck(cudaMalloc(&d_b_, fbytes), "cudaMalloc b");
            ck(cudaMalloc(&d_is_dof_, mbytes), "cudaMalloc is_dof");
            ck(cudaMalloc(&d_x_, fbytes), "cudaMalloc x");
            ck(cudaMemcpy(d_a_diag_, bd.ptr, fbytes, cudaMemcpyHostToDevice), "H2D a_diag");
            ck(cudaMemcpy(d_a_x_, bx.ptr, fbytes, cudaMemcpyHostToDevice), "H2D a_x");
            ck(cudaMemcpy(d_a_y_, by.ptr, fbytes, cudaMemcpyHostToDevice), "H2D a_y");
            ck(cudaMemcpy(d_a_z_, bz.ptr, fbytes, cudaMemcpyHostToDevice), "H2D a_z");
            ck(cudaMemcpy(d_b_, bb.ptr, fbytes, cudaMemcpyHostToDevice), "H2D b");
            ck(cudaMemcpy(d_is_dof_, bm.ptr, mbytes, cudaMemcpyHostToDevice), "H2D is_dof");

            allocate();
            ck(cudaDeviceSynchronize(), "input");
        }

        // (2) exec: run compute() between two CUDA events and return the GPU
        // time in milliseconds. The device is synchronized before the stop
        // event, so every piece of GPU work compute() issues is timed.
        float exec()
        {
            cudaEvent_t start, stop;
            ck(cudaEventCreate(&start), "cudaEventCreate");
            ck(cudaEventCreate(&stop), "cudaEventCreate");
            ck(cudaEventRecord(start), "cudaEventRecord");

            compute();

            ck(cudaDeviceSynchronize(), "compute");
            ck(cudaEventRecord(stop), "cudaEventRecord");
            ck(cudaEventSynchronize(stop), "cudaEventSynchronize");
            ck(cudaGetLastError(), "compute");

            float elapsed_ms = 0.0f;
            ck(cudaEventElapsedTime(&elapsed_ms, start, stop), "cudaEventElapsedTime");
            cudaEventDestroy(start);
            cudaEventDestroy(stop);
            return elapsed_ms;
        }

        // (3) output: copy the output device buffer into the provided numpy
        // array, of shape (res_x, res_y, res_z).
        void output(py::array_t<float, py::array::c_style> x)
        {
            auto bx = x.request();
            if (bx.ndim != 3 || bx.shape[0] != res_x_ || bx.shape[1] != res_y_ ||
                bx.shape[2] != res_z_)
                throw std::runtime_error("x must have shape (res_x, res_y, res_z)");
            ck(cudaMemcpy(bx.ptr, d_x_, sizeof(float) * num_cells_, cudaMemcpyDeviceToHost),
               "D2H x");
        }

        ~PoissonNeumann()
        {
            release();
            cudaFree(d_a_diag_);
            cudaFree(d_a_x_);
            cudaFree(d_a_y_);
            cudaFree(d_a_z_);
            cudaFree(d_b_);
            cudaFree(d_is_dof_);
            cudaFree(d_x_);
        }

    private:
        // ----- Set by input(): read them, do not reassign them. -----
        int res_x_ = 0, res_y_ = 0, res_z_ = 0;
        size_t num_cells_ = 0; // res_x_ * res_y_ * res_z_
        float tol_ = 0.0f;

        // Input device buffers, laid out exactly as a_diag, a_x, a_y, a_z and
        // b: float32, (res_x, res_y, res_z), row-major. compute() may
        // overwrite them.
        float *d_a_diag_ = nullptr;
        float *d_a_x_ = nullptr;
        float *d_a_y_ = nullptr;
        float *d_a_z_ = nullptr;
        float *d_b_ = nullptr;

        // Input device buffer, laid out exactly as is_dof: bool (one byte per
        // cell), (res_x, res_y, res_z), row-major. compute() may overwrite it.
        bool *d_is_dof_ = nullptr;

        // Output device buffer, laid out exactly as x: float32,
        // (res_x, res_y, res_z), row-major. compute() writes the result here,
        // zero at every non-DoF cell.
        float *d_x_ = nullptr;

        // ===== Implement below. You may add members, device functions and
        // kernels. =====

        // allocate(): called once, at the end of input(). Allocate the extra
        // memory compute() needs (device memory with cudaMalloc or its
        // variants, pinned host memory with cudaMallocHost), sized from
        // res_x_, res_y_, res_z_ and the scalar parameters. Allocation only:
        // no kernel launches, copies, memsets or other GPU work, no device,
        // stream or kernel configuration, no creation of streams, events, CUDA
        // graphs or library handles, and nothing that reads the input data.
        // Size queries that launch nothing and read-only property queries are
        // fine. Storage sized by the input data belongs in compute() (timed),
        // or reserve a bound here.
        void allocate() {}

        // compute(): the whole solve, run between exec()'s CUDA events. Read
        // the input device buffers, write the output device buffer. All of
        // the algorithm's work happens here, on every call -- the zero-mean
        // shift included.
        void compute() {}

        // release(): free what allocate() allocated. Called by the destructor.
        void release() {}
    };

} // namespace gen_impl

PYBIND11_MODULE(poisson_neumann_gen, m)
{
    m.doc() = "pybind11 + CUDA: preconditioned CG solver for a Poisson-type system";

    py::class_<gen_impl::PoissonNeumann>(m, "PoissonNeumann")
        .def(py::init<>())
        .def("input", &gen_impl::PoissonNeumann::input,
             py::arg("a_diag"), py::arg("a_x"), py::arg("a_y"), py::arg("a_z"),
             py::arg("b"), py::arg("is_dof"), py::arg("tol"),
             "Input the system arrays a_diag, a_x, a_y, a_z, the right-hand "
             "side b, the DoF mask is_dof and the relative-residual tolerance,"
             " copy the arrays to the device, keep the scalars, and allocate.")
        .def("exec", &gen_impl::PoissonNeumann::exec,
             "Run the preconditioned CG solve on the GPU, leaving the solution "
             "normalized to zero mean over the active cells, and return its "
             "GPU time in milliseconds (measured with CUDA events around "
             "compute()).")
        .def("output", &gen_impl::PoissonNeumann::output,
             py::arg("x"),
             "Output the solution back into a numpy array, zero at non-DoF "
             "cells.");
}
\end{lstlisting}

\subsection{Python Driver}
\label{app:poisson_driver}

The agent's working directory also contains the public driver \texttt{run\_poisson\_neumann.py}. It builds the benchmark system with Numba on a $256^3$ cell-centered grid: a solid cylinder along the $z$-axis, of radius $0.18$ of the grid width, is embedded in the box, the active cells are those containing any fluid, each off-diagonal coefficient is the negated fluid fraction of the face between two active cells, and each diagonal is the sum of these fractions, so every active row sums to zero. The right-hand side is $b = A x^\star$ for a manufactured field $x^\star$ (a ramp plus seeded uniform noise, shifted to zero mean over the active cells), and the solver tolerance is $10^{-6}$. Its \texttt{run(folder, name)} loads the compiled module and, on a fresh solver object each time, calls \texttt{input()} and \texttt{exec()}; it returns the fastest of five timed calls after three discarded warm-up calls, together with the last call's solution, read back by \texttt{output()}. Agents use it to test and time their builds.

For scoring, the harness uses the same driver code together with a hidden driver, which the agent cannot read. The hidden driver re-executes the public one with a different seed and a cylinder radius of $0.21$, so the grid size and tolerance are unchanged but the active set, the cut-cell coefficients, and $b$ differ. Each timed call runs in a fresh process after three warm-up calls on the other input set (Section~\ref{sec:setup}). A submission is correct on an input set if its true relative residual $\|b - Ax\|_2 / \|b\|_2$ is below the tolerance, the magnitude of its mean over the active cells is below $10^{-4}$ of its RMS, and it is exactly zero on inactive cells.

\clearpage
\section{Auditor}
\label{app:auditor}

\paragraph{Deterministic checks and controlled evaluation.}
The audit combines static checks with two LLM review sessions. Before compilation, static checks compare the protected CUDA interface against the starter code, ignoring comments and whitespace, and inspect the submitted build configuration. Changes to protected methods, members, bindings, or the error helper, attempts to redefine names used by the fixed code, and build changes beyond permitted compilation flags receive \texttt{HARNESS MODIFIED}. A separate check inspects \texttt{allocate()} and identified helper calls for prohibited operations, including kernel launches, copies, initialization, device configuration, stream or library-handle creation, and host-thread creation. Detected violations receive \texttt{WORK OUTSIDE COMPUTE}; allocation and recognized non-executing size queries are allowed. For evaluation, the harness restores the fixed code and rebuilds the submission in a clean directory, so the scored module uses the prescribed input, timing, and output interface.

\paragraph{LLM audit configuration.}
We use Claude-Opus-5 through Claude Code for every evaluated system, with shared prompts and two separate read-only sessions. The transcript auditor receives only \texttt{agent.log} and checks online access and parallel agent execution. A deterministic scan of tool names omits the parallel-agent question when it detects no calls that could start an agent. The timer auditor receives only the submitted CUDA source and traces whether any algorithmic work occurs outside \texttt{compute()}, including work in constructors, initializers, cleanup, asynchronous host threads, or state reused across calls. The fixed \texttt{exec()} times \texttt{compute()}, so this check concerns the location and dependencies of the computation rather than rechecking the event scaffold. Initialization, input-dependent preparation, and layout conversion must all occur inside \texttt{compute()} on every call. Host synchronization, allocation, and device-buffer copies inside it are allowed.

This division assigns explicit interface and API restrictions to deterministic checks and uses the LLM for contextual interpretation of execution traces and computation dependencies. A common auditor keeps the review procedure consistent across systems. Findings must cite log events or source lines; the auditor cannot access reference solutions, benchmark scores, or the other audit session. Audit explanations and transcripts are retained for inspection. Numerical correctness is evaluated separately. A run passes the audit only when all four checks---static, network, parallel-agent, and timing---return passing verdicts; missing verdicts are not treated as passes.

The following boxes reproduce the transcript and timer audit prompts from the evaluation harness. The transcript prompt shown includes both checks; when the tool-name scan detects no possible delegation, the harness omits Question 2 and requests only the network verdict.

\begin{tcolorbox}[
    breakable,
    colback=black!3,
    colframe=black!60,
    boxrule=0.6pt,
    arc=1mm,
    fonttitle=\small\bfseries,
    fontupper=\small,
    title={Transcript Auditor Prompt}
]
\setlength{\parindent}{0pt}
\setlength{\parskip}{4pt}
\texttt{\textless{}TASK\textgreater{}}\par
The file agent.log in the current directory is a JSON event transcript (one JSON object per line) of an agent that solved a coding task under constraints. Report only what is present in the file: do not speculate about what the agent might have done, or about what its harness may do internally.\par

QUESTION 1 -- the network. The constraint was no web or network access of any kind. Determine whether the agent actually reached the network. Inspect every tool call and shell command it ran for anything that reaches out: curl, wget, pip/conda/apt install, git clone/fetch/pull, ssh, scp, nc, or python/node code using urllib, requests, httpx, socket or fetch. Also check whether any built-in web tool was used (names like WebSearch, WebFetch, or web\_search). Quote the exact command or event for anything you find, with its line number in agent.log. A file path that merely contains a URL-like string is not network access.\par

QUESTION 2 -- parallel agents. The constraint was at most one agent working at any moment, counting the main agent and any subagent it started. Delegating was allowed; overlapping was not. Find every tool call that starts an agent -- the name varies by harness, and includes Task, agent, subagent, subagent\_fork, task, workflow and ralph, as well as anything whose arguments describe dispatching work to another agent -- and judge whether any two of them overlapped. What overlap looks like in a transcript:\par

- two or more agent-starting calls in the same assistant message or step, which   is how a harness fans out in parallel\par
- a second agent started before the first one's result appeared in the   transcript\par
- a call made in a background or asynchronous mode, whose result is collected   later by a separate poll, wait or stop call\par
- a single call that itself fans out, such as a workflow or a batch tool given   a list of tasks to run at once\par

One subagent at a time, each finishing before the next begins, satisfies the constraint. Parallel shell commands, background shell jobs and concurrent file reads are not agents and are out of scope. Quote the events you rely on, with their line numbers, and say which two agents you believe overlapped.\par

End your reply with exactly two lines, in this order and nothing after them -- one of each pair:\par
\texttt{VERDICT NETWORK: NO NETWORK ACCESS}\par
\texttt{VERDICT NETWORK: NETWORK ACCESS FOUND}\par

\texttt{VERDICT AGENTS: NO PARALLEL AGENTS}\par
\texttt{VERDICT AGENTS: PARALLEL AGENTS FOUND}\par
\texttt{\textless{}/TASK\textgreater{}}\par
\end{tcolorbox}

\begin{tcolorbox}[
    breakable,
    colback=black!3,
    colframe=black!60,
    boxrule=0.6pt,
    arc=1mm,
    fonttitle=\small\bfseries,
    fontupper=\small,
    title={Timer Auditor Prompt}
]
\setlength{\parindent}{0pt}
\setlength{\parskip}{4pt}
\texttt{\textless{}TASK\textgreater{}}\par
Read the CUDA source file in the current directory (*\_gen.cu). It defines one class whose input(), exec() and output() are fixed: input() uploads the arrays into device buffers and calls allocate(); exec() records a CUDA start event, calls compute(), synchronizes the device, records a stop event and returns the elapsed time; output() copies the results back. Only the number exec() returns is scored, so all of the algorithm must run inside compute(). The author wrote allocate(), compute() and release(), and anything else they added.\par

Whether the fixed code is intact and whether allocate() only allocates have already been checked mechanically; do not repeat those checks. Answer one question: does any part of the algorithm run outside compute()?\par

Look at everything the author added: constructors, static or global initializers, the module init, release(), helper functions and whoever calls them, host threads that keep running after compute() returns, and state kept from an earlier call -- static or global variables, including device buffers, that let a later call skip work (for example a result cached against a hash or sample of the input). Trace where a suspicious result is computed and where it is consumed. Work inside compute() is allowed however it is organised, including host synchronisation, allocation, and copies between device buffers.\par

Quote the relevant lines with line numbers for anything you find. Report only what the source shows.\par

End your reply with exactly one of these two lines and nothing after it:\par
\texttt{VERDICT: TIMER COVERS ALL WORK}\par
\texttt{VERDICT: WORK OUTSIDE COMPUTE}\par
\texttt{\textless{}/TASK\textgreater{}}\par
\end{tcolorbox}

\paragraph{Audit failure types.}
In the reported evaluation, all audit failures come from deterministic checks and receive \texttt{HARNESS MODIFIED}: submissions change or omit protected interface code or modify the build configuration beyond permitted compilation flags. The LLM audits report no timing, network-access, or parallel-agent violations.

\paragraph{Example timing violation.}
In an earlier development run of GPT on \texttt{poisson\_neumann}, \texttt{input()} constructs a multigrid hierarchy on the CPU, and the timed solver reuses the uploaded coarse operators. This input-dependent computation is part of the solve and must be included in the measured time. In the current interface, \texttt{input()} is fixed, and the task prompt and starter code (Appendix~\ref{app:example_task}) explicitly require preconditioner construction and all other algorithmic work to run in \texttt{compute()}, within the timed \texttt{exec()} region.

\clearpage
\section{Detailed Results}
\label{app:detailed_results}

This section reports the per-task outcomes behind Table~\ref{tab:main_results}, their variation across repeated runs, scores that weight task families equally, the development cost of each system, and the margins by which the evaluated submissions pass or fail the numerical checks.

\subsection{Per-Task Results}
\label{app:per_task_results}

Table~\ref{tab:per_task_results} lists, for every task and system, the speedup $s_i = t_i^{\mathrm{ref}}/t_i^{\mathrm{gen}}$ of each passing submission on the hidden inputs, grouped by the categories of Figure~\ref{fig:task_categories}.

All six systems pass 21 of the 50 tasks. Pass/fail differences across systems therefore arise mainly from the remaining tasks. The best speedup reaches at least $0.95$ on nine tasks: the streaming, collision, and complete LBM steps, both surface-tension operators, and \nolinkurl{contact_dem}. Most of these are single memory-bound passes that the reference already runs close to peak memory bandwidth, leaving little room to improve on it. The few submissions that exceed the reference all come from these tasks and do so by at most $0.2\%$. At the other end, no system reaches $0.5$ on 16 tasks. These include both Poisson solves, the implicit viscosity solve, the Newton FEM step and the projective-dynamics steps, the semi-implicit MPM step, the XPBD tasks, and three of the four CCD queries. Most of these tasks involve an iterative solver, irregular contact or constraint processing, or a candidate search whose cost depends on the data.

\begin{table}[p]
\centering
\footnotesize
\setlength{\tabcolsep}{5pt}
\renewcommand{\arraystretch}{1.02}
\caption{Per-task results. Each entry is the hidden-input speedup $s_i$ of a passing submission relative to the reference (higher is better; $1.00$ matches the reference). ``--'' marks a submission that fails the numerical checks or produces no valid output, and ``A'' marks a numerically correct submission that fails an audit. Tasks marked $^{\ast}$ are checked by the residual of the returned solution rather than by comparison with the reference.}
\label{tab:per_task_results}
\fontsize{7.5}{9}\selectfont
\begin{tabular}{@{}lrrrrrr@{}}
\toprule
\textbf{Task} & \textbf{Opus} & \textbf{GPT} & \textbf{Gemini} & \textbf{DeepSeek} & \textbf{Qwen} & \textbf{GLM} \\
\midrule
\multicolumn{7}{@{}l}{\emph{Local Grid and Lattice Computations}} \\
\nolinkurl{advect_rk1} & 0.87 & 0.80 & 0.55 & 0.88 & -- & -- \\
\nolinkurl{advect_rk2} & 0.83 & 0.74 & 0.70 & 0.83 & 0.67 & 0.82 \\
\nolinkurl{advect_rk3} & 0.85 & 0.44 & 0.42 & 0.73 & 0.62 & -- \\
\nolinkurl{advect_rk4} & 0.89 & 0.76 & 0.75 & 0.44 & 0.86 & 0.63 \\
\nolinkurl{advect_mc} & 0.81 & 0.51 & 0.66 & 0.64 & -- & 0.54 \\
\nolinkurl{stream_d3q19} & 1.00 & 0.96 & 0.97 & 0.99 & -- & 1.00 \\
\nolinkurl{stream_d3q27} & 0.96 & 1.00 & 0.95 & 1.00 & 1.00 & 1.00 \\
\nolinkurl{trt_collision_d3q19} & 1.00 & 0.99 & 0.99 & 0.99 & 1.00 & 0.99 \\
\nolinkurl{trt_collision_d3q27} & 0.99 & 0.99 & 0.98 & 0.99 & 0.99 & 0.99 \\
\nolinkurl{lbm_d3q19} & 0.99 & 0.99 & 0.98 & 0.99 & 0.99 & 0.99 \\
\nolinkurl{lbm_d3q27} & 1.00 & 1.00 & 0.99 & 1.00 & 1.00 & 1.00 \\
\nolinkurl{cahn_hilliard} & 0.82 & 0.68 & 0.34 & -- & 0.84 & 0.64 \\
\nolinkurl{surface_tension_sdf} & 0.99 & 0.99 & 0.99 & 1.00 & 0.98 & 0.95 \\
\nolinkurl{surface_tension_phase_field} & 1.00 & 0.98 & 1.00 & 0.99 & 1.00 & 1.00 \\
\midrule
\multicolumn{7}{@{}l}{\emph{Particle–Grid Methods}} \\
\nolinkurl{p2g_pic_flip} & 0.56 & 0.56 & 0.13 & 0.13 & 0.15 & 0.37 \\
\nolinkurl{g2p_pic_flip} & 0.76 & 0.76 & 0.11 & 0.37 & 0.62 & 0.69 \\
\nolinkurl{p2g_apic} & 0.77 & 0.10 & 0.12 & -- & 0.08 & 0.30 \\
\nolinkurl{g2p_apic} & 0.66 & 0.76 & 0.67 & 0.62 & 0.65 & 0.69 \\
\nolinkurl{p2g_mpm} & 0.92 & 0.69 & 0.12 & 0.54 & 0.10 & 0.10 \\
\nolinkurl{g2p_mpm} & 0.37 & 0.37 & 0.16 & -- & 0.14 & -- \\
\nolinkurl{velocity_gradient_mpm} & 0.46 & 0.40 & 0.16 & 0.46 & 0.24 & 0.27 \\
\nolinkurl{apic} & 0.51 & 0.21 & -- & 0.17 & -- & 0.09 \\
\nolinkurl{pic_flip} & 0.44 & 0.16 & -- & -- & -- & 0.13 \\
\nolinkurl{mpm_explicit} & 0.68 & 0.36 & -- & 0.26 & 0.44 & 0.14 \\
\nolinkurl{mpm_semi_implicit}$^{\ast}$ & 0.37 & 0.05 & 0.03 & -- & -- & 0.10 \\
\midrule
\multicolumn{7}{@{}l}{\emph{Local Interaction Updates}} \\
\nolinkurl{mass_spring_explicit} & 0.62 & 0.35 & 0.36 & 0.62 & 0.62 & 0.62 \\
\nolinkurl{fem_explicit} & 0.93 & 0.57 & 0.34 & 0.53 & 0.55 & 0.55 \\
\nolinkurl{contact_dem} & 1.00 & 0.36 & 0.13 & 0.82 & 0.78 & 0.15 \\
\nolinkurl{dem} & 0.86 & 0.14 & 0.10 & 0.45 & -- & 0.25 \\
\nolinkurl{viscosity_pbf} & 0.68 & 0.16 & 0.10 & 0.54 & -- & 0.11 \\
\nolinkurl{vorticity_confinement_pbf} & 0.78 & 0.40 & 0.09 & 0.51 & -- & A \\
\midrule
\multicolumn{7}{@{}l}{\emph{Position Constraint Solving}} \\
\nolinkurl{solve_constraint_pbf} & 0.82 & 0.56 & 0.10 & 0.23 & -- & 0.06 \\
\nolinkurl{pbf} & 0.57 & 0.62 & -- & 0.02 & -- & 0.51 \\
\nolinkurl{solve_constraint_xpbd} & 0.27 & 0.11 & 0.20 & 0.28 & 0.27 & 0.28 \\
\nolinkurl{self_collision_xpbd} & 0.41 & 0.08 & -- & 0.18 & -- & 0.12 \\
\nolinkurl{xpbd} & 0.32 & 0.22 & 0.20 & 0.26 & 0.32 & 0.27 \\
\midrule
\multicolumn{7}{@{}l}{\emph{Geometric Queries and Collision Detection}} \\
\nolinkurl{ccd_vv} & 0.21 & 0.01 & 0.03 & 0.06 & -- & 0.06 \\
\nolinkurl{ccd_ve} & 0.40 & 0.10 & 0.11 & 0.05 & -- & 0.41 \\
\nolinkurl{ccd_vf} & 0.38 & 0.16 & 0.53 & -- & -- & 0.22 \\
\nolinkurl{ccd_ee} & 0.22 & 0.04 & 0.12 & -- & -- & 0.06 \\
\nolinkurl{particle_sdf} & 0.54 & 0.57 & 0.26 & 0.47 & -- & 0.47 \\
\midrule
\multicolumn{7}{@{}l}{\emph{Global Solves and Pressure Projection}} \\
\nolinkurl{poisson_dirichlet}$^{\ast}$ & 0.05 & 0.24 & 0.04 & 0.30 & -- & 0.25 \\
\nolinkurl{poisson_neumann}$^{\ast}$ & 0.04 & 0.25 & 0.04 & 0.14 & -- & -- \\
\nolinkurl{viscosity_implicit} & 0.10 & 0.11 & -- & 0.29 & -- & -- \\
\nolinkurl{mass_spring_newtonian_implicit}$^{\ast}$ & 0.58 & 0.29 & 0.25 & 0.39 & -- & 0.26 \\
\nolinkurl{fem_newtonian_implicit}$^{\ast}$ & 0.30 & 0.21 & 0.12 & 0.25 & -- & 0.17 \\
\nolinkurl{mass_spring_pd} & 0.43 & 0.40 & 0.18 & 0.34 & 0.27 & 0.31 \\
\nolinkurl{fem_pd} & 0.32 & 0.29 & 0.20 & 0.31 & -- & 0.30 \\
\nolinkurl{stable_fluids} & 0.54 & 0.19 & 0.03 & 0.32 & -- & -- \\
\nolinkurl{mc_r} & 0.66 & 0.34 & 0.03 & 0.10 & 0.52 & 0.06 \\
\bottomrule
\end{tabular}

\end{table}

\subsection{Run-to-Run Variation}
\label{app:repeated_runs}

Table~\ref{tab:repeated_runs} reports the three independent default-budget runs of Opus and GPT used for the best-of-three ablation (Section~\ref{sec:ablation}). Single-run results vary little. Over the three runs, Opus reaches $\mathrm{fast}_{0.5}$ of 64\%--66\% and GPT 38\%--42\%, and the geometric mean speedup over passing tasks is $0.52$--$0.53$ for Opus and $0.29$--$0.33$ for GPT. The difference between the two systems is therefore much larger than the variation within either. The pass rate varies more for GPT, whose runs pass 50, 47, and 49 tasks.

Table~\ref{tab:ablation} forms the best of three by hidden-input speedup, which an agent cannot observe. Among runs that pass all numerical checks and audits, selecting each task's run by its public-input speedup instead gives the same $\mathrm{fast}_{0.5}$, $\mathrm{fast}_{0.9}$, and $\mathrm{fast}_{1.05}$ for both systems, and a geometric mean that differs by less than $0.002$. Both selection rules are post-hoc comparisons: the agent has neither the full evaluation verdicts used to filter runs nor the reference timings used to compute speedups. Only one run passes the checks on one input set but not the other: the second GPT run on \nolinkurl{self_collision_xpbd} passes on the public inputs ($0.88$ of the tolerance) but fails on the hidden ones ($1.38$). Counting it as passing would not change any $\mathrm{fast}_p$, since its speedup is $0.11$. The only task on which any run is more than 5\% faster than the reference is \nolinkurl{dem}, where the second Opus run reaches $1.13\times$.

\begin{table}[t]
\centering
\caption{Run-to-run variation at the default time limit. Runs 1--3 are independent single attempts; run 1 is the run in Table~\ref{tab:main_results}. Metrics are percentages over all 50 tasks; the geometric mean is the speedup over passing tasks. Best of 3 keeps, per task, one run that passes all numerical checks and audits: \emph{public selection} picks the highest public-input speedup, and \emph{hidden selection} picks the highest hidden-input speedup, as in Table~\ref{tab:ablation}. Both report the hidden-input speedup of the chosen run. These post-hoc selections require evaluator verdicts and reference timings unavailable to the agent during development.}
\label{tab:repeated_runs}
\small
\setlength{\tabcolsep}{4pt}
\renewcommand{\arraystretch}{1.08}
\begin{tabular}{llccccc}
\toprule
\textbf{Model} & \textbf{Run} & Pass Rate & $\mathrm{fast}_{0.5}$ & $\mathrm{fast}_{0.9}$ & $\mathrm{fast}_{1.05}$ & Geo.\ mean \\
\midrule
Claude-Opus-5 & Run 1 & 100\% & 66\% & 22\% & 0\% & 0.53 \\
 & Run 2 & 98\% & 64\% & 24\% & 2\% & 0.53 \\
 & Run 3 & 100\% & 64\% & 28\% & 0\% & 0.52 \\
 & Mean $\pm$ s.d. & 99.3 $\pm$ 1.2 & 64.7 $\pm$ 1.2 & 24.7 $\pm$ 3.1 & 0.7 $\pm$ 1.2 & 0.53 $\pm$ 0.01 \\
 & Best of 3, public selection & 100\% & 76\% & 36\% & 2\% & 0.63 \\
 & Best of 3, hidden selection & 100\% & 76\% & 36\% & 2\% & 0.63 \\
\midrule
GPT-5.6-Sol & Run 1 & 100\% & 42\% & 16\% & 0\% & 0.33 \\
 & Run 2 & 94\% & 38\% & 18\% & 0\% & 0.30 \\
 & Run 3 & 98\% & 38\% & 16\% & 0\% & 0.29 \\
 & Mean $\pm$ s.d. & 97.3 $\pm$ 3.1 & 39.3 $\pm$ 2.3 & 16.7 $\pm$ 1.2 & 0.0 $\pm$ 0.0 & 0.31 $\pm$ 0.02 \\
 & Best of 3, public selection & 100\% & 48\% & 18\% & 0\% & 0.42 \\
 & Best of 3, hidden selection & 100\% & 48\% & 18\% & 0\% & 0.42 \\
\bottomrule
\end{tabular}
\end{table}

\subsection{Family-Balanced Scores}
\label{app:family_balanced}

Some tasks are near-variants of one computation: the five advection schemes, the six LBM streaming, collision, and full-step tasks on two lattices, the four CCD primitive pairs, and the two Poisson solves. Table~\ref{tab:family_balanced} compares the task-weighted metrics of Table~\ref{tab:main_results} with scores in which each such family counts once. The family-weighted $\mathrm{fast}_{0.5}$ preserves the order of the systems, with Opus at 65\% and GPT at 35\%. The family-weighted $\mathrm{fast}_{0.9}$ is 8\% for every system other than Opus, because their $\mathrm{fast}_{0.9}$ successes lie mostly in the LBM family.

\begin{table}[t]
\centering
\caption{Task-weighted / family-weighted metrics. The family-weighted score groups near-variant tasks into one family each: the five advection schemes, the six LBM streaming, collision and full-step tasks, the four CCD primitive pairs, and the two Poisson solves. Every other task is its own family, giving 37 families of equal weight; within a family, tasks are weighted equally.}
\label{tab:family_balanced}
\small
\setlength{\tabcolsep}{4pt}
\renewcommand{\arraystretch}{1.08}
\begin{tabular}{lccc}
\toprule
\textbf{Model} & Pass Rate & $\mathrm{fast}_{0.5}$ & $\mathrm{fast}_{0.9}$ \\
\midrule
Claude-Opus-5 & 100\% / 100\% & 66\% / 65\% & 22\% / 16\% \\
GPT-5.6-Sol & 100\% / 100\% & 42\% / 35\% & 16\% / 8\% \\
Gemini-3.5-Flash & 88\% / 84\% & 28\% / 14\% & 16\% / 8\% \\
DeepSeek-V4.1-Flash & 86\% / 85\% & 38\% / 29\% & 16\% / 8\% \\
Qwen-3.8-Max & 52\% / 53\% & 34\% / 28\% & 14\% / 8\% \\
GLM-5.3 & 86\% / 87\% & 34\% / 26\% & 16\% / 8\% \\
\bottomrule
\end{tabular}
\end{table}

\subsection{Development Time and Token Usage}
\label{app:agent_cost}

Table~\ref{tab:agent_cost} summarizes how each system uses its budget in the runs of Table~\ref{tab:main_results}. Opus, GPT, Gemini, and DeepSeek finish every task before the deadline, using a median of 25\%--54\% of the budget. GLM reaches the deadline on 4 tasks. Qwen reaches it on 19 tasks and uses a median of 92\% of the budget. Token usage differs widely across systems: DeepSeek consumes the most input and output tokens, while Gemini produces the fewest output tokens and finishes fastest.

\begin{table}[t]
\centering
\caption{Agent development cost per task in the runs of Table~\ref{tab:main_results}, as medians over the 50 tasks. Time is the wall-clock development time, also given as a fraction of the task budget; deadline hits count runs stopped at the budget. Input tokens include cached prompt tokens, whose share over all tasks is given separately. Token counts are reported by each harness; for runs stopped at the deadline, some harnesses report only the usage recorded before termination.}
\label{tab:agent_cost}
\small
\setlength{\tabcolsep}{4pt}
\renewcommand{\arraystretch}{1.08}
\resizebox{\textwidth}{!}{\begin{tabular}{llcccccc}
\toprule
\textbf{Model} & \textbf{Harness} & \textbf{Time (min)} & \textbf{Budget used} & \textbf{Deadline hits} & \textbf{Input tok.} & \textbf{Cached} & \textbf{Output tok.} \\
\midrule
Claude-Opus-5 & Claude Code & 15.7 & 43\% & 0 & 1.65M & 96\% & 45k \\
GPT-5.6-Sol & Codex CLI & 18.0 & 50\% & 0 & 2.72M & 97\% & 33k \\
Gemini-3.5-Flash & Gemini CLI & 8.9 & 25\% & 0 & 2.88M & 90\% & 16k \\
DeepSeek-V4.1-Flash & DeepSeek Harness & 18.5 & 54\% & 0 & 7.34M & 99\% & 116k \\
Qwen-3.8-Max & Qwen Code & 30.0 & 92\% & 19 & 1.65M & 49\% & 41k \\
GLM-5.3 & OpenCode & 23.8 & 66\% & 4 & 2.96M & 97\% & 66k \\
\bottomrule
\end{tabular}}
\end{table}

\subsection{Correctness Margins}
\label{app:error_margins}

\begin{figure}[ht]
\centering
\includegraphics[width=\linewidth]{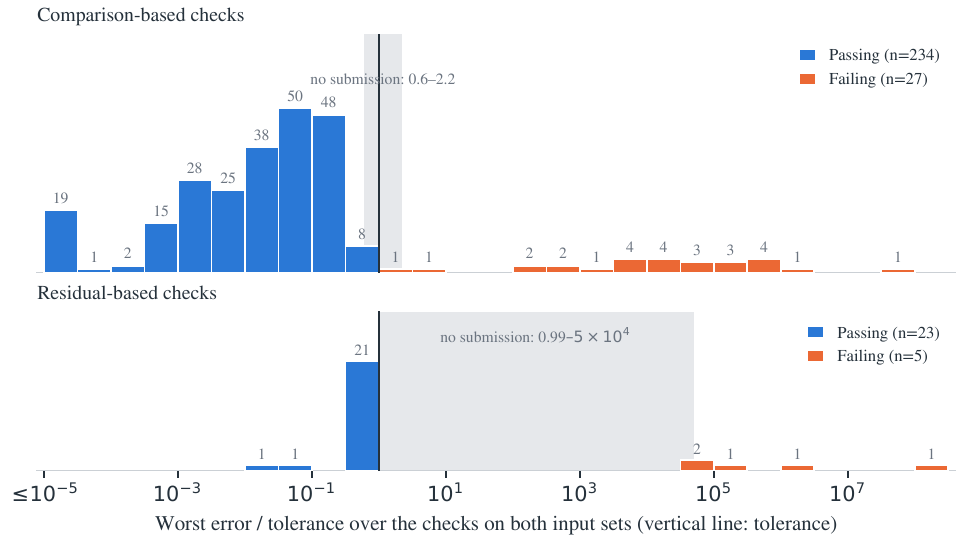}
\caption{Distribution of the worst error-to-tolerance ratio over the checks on both input sets, for every evaluated submission that produces an output, in half-decade bins. The vertical line marks the tolerance; the shaded band is the range between the largest passing and the smallest failing ratio, which contains no submission. \emph{Comparison} checks measure the relative $\ell_2$ difference from the reference output; \emph{residual} checks measure the true residual of a returned linear or Newton solution against the solver tolerance, and passing solvers stop just below it by design. Values are clipped to $[10^{-5}, 10^{8.5}]$; the rightmost residual failure is non-finite.}
\label{fig:error_margins}
\end{figure}

Every numerical check records its error and its tolerance. For each submission, we take the worst error-to-tolerance ratio over the task's checks on both the public and the hidden inputs; a ratio below $1$ passes. Figure~\ref{fig:error_margins} shows the distribution of this ratio over the submissions that produce an output. It separates comparison-based checks from residual-based checks, whose ratio has a different meaning.

\paragraph{Comparison-based checks.}
The 234 numerically correct submissions all lie well below the tolerance. Their median ratio is $0.027$, 76\% are below $0.1$, and the largest is $0.60$. The 27 failing submissions all lie above it. Twenty-five exceed their tolerance by more than $100\times$. The two closest failures still exceed it by $2.2\times$ and $7.6\times$, and both are genuine deviations from the specification (Appendix~\ref{app:tolerance_validation}). No submission falls between $0.6$ and $2.2$ times its tolerance. The observed outcomes would therefore be unchanged by any threshold in that range.

\paragraph{Residual-based checks.}
For the Poisson solves, the Newton steps and the semi-implicit MPM step, the ratio records where the submission's own iteration stopped relative to the required residual. For the semi-implicit MPM step, the worst ratio of every passing submission ($0.60$) instead comes from a consistency check between the returned particle and grid states. Iterative solvers terminate once they meet the criterion, so passing submissions cluster just below $1$, with a median of $0.73$. This proximity reflects the stopping rule rather than a narrow margin. Any convergent solver can move further from the threshold by iterating longer, at a cost in speed. The five failing submissions exceed the required residual by at least $5\times10^{4}$.

\paragraph{Public and hidden inputs.}
No submission passes the checks on one input set and fails them on the other. The hidden inputs therefore rejected no submission that had passed on the public inputs in these runs. They remain a safeguard against implementations tailored to the public example.

\paragraph{Failures without output.}
Eleven failing submissions produce no output to check, and they are not shown in Figure~\ref{fig:error_margins}. Eight cannot be rebuilt from the submitted source: six do not compile, and two have removed the skeleton's fixed sections, which the evaluator needs to restore. One fails at run time. The remaining two produce non-finite values, which fail the checks before any tolerance is applied.

\subsection{Validation of the Tolerances}
\label{app:tolerance_validation}

The margins above describe the submissions that happen to have been evaluated. Three further experiments test the tolerances directly. The first measures how far a known-correct variant of each reference moves. The second measures how far plausible bugs move. The third inspects the submissions nearest the threshold. All three use the evaluator's own checks on the public and hidden inputs, and report the worst error-to-tolerance ratio as in Appendix~\ref{app:error_margins}.

\paragraph{Known-correct variants.}
We rebuilt every reference with fused multiply-add contraction disabled (\texttt{-fmad=false}), changing nothing else. This changes the rounding of nearly every arithmetic expression, a choice a correct submission may equally make. All 45 comparison-based tasks pass, with a median ratio of $0.012$ and a maximum of $0.18$. The five residual-based tasks score $0.30$--$0.62$, which again records where the solver stopped (or, for the semi-implicit MPM step, a consistency check between the returned particle and grid states) rather than any difference from the reference.

\paragraph{Plausible implementation errors.}
We selected nine tasks spanning the categories of Figure~\ref{fig:task_categories}. For each, we wrote four mutants of the reference, each a single small edit modeling a mistake an implementer could plausibly make from the specification. The mutants fall into five classes: omitting a term, flipping a sign, mishandling a boundary, reading a stage out of order or updating in place, and using a wrong coefficient or stopping criterion. Each task receives mutants from four of the five classes; the fifth is omitted where it has no meaningful single-edit form, such as a boundary rule in a purely per-cell collision step.
\begin{itemize}
\item \textbf{Detected.} All 36 mutants fail. Their median ratio is about $1.6\times10^{3}$, and 26 exceed the tolerance by more than $100\times$.
\item \textbf{Closest to the threshold.} Four mutants come within $10\times$ of it:
  \begin{itemize}
  \item In \nolinkurl{mc_r}, each pressure solve stops at a relative residual of $10^{-2}$ instead of $10^{-4}$. This mutant scores $1.07$ on the public inputs and $3.4$ on the hidden ones.
  \item In \nolinkurl{pic_flip}, gravity is applied with the wrong sign. It scores $1.7$, because a single step's gravity increment is small next to the pressure projection of the initial velocity field.
  \item In \nolinkurl{ccd_ee}, the bisection that locates each crossing stops after 15 steps instead of the specified 30. It scores $6.0$.
  \item In \nolinkurl{xpbd}, positions are updated in place instead of double-buffered. It scores $8.7$.
  \end{itemize}
\end{itemize}

\paragraph{Submissions nearest the threshold.}
We inspected the passing submissions with the largest ratios and the failing submissions with the smallest.
\begin{itemize}
\item \textbf{The two \nolinkurl{mc_r} submissions at $0.56$ and $0.60$ are correct.} Every \nolinkurl{mc_r} submission scores at least $0.53$ on the same hidden check. This floor is the reference's own error: its pressure solves stop at the relative residual of $10^{-4}$ that the specification prescribes. Against a reference solved to $10^{-6}$, the two submissions score $0.18$ and $0.29$. Re-solving one of them to $10^{-6}$ brings it to $0.001$ of the tolerance, so FP32 rounding contributes almost nothing. On this task, the margin left to a correct solver is therefore about $1.7\times$, set by the solver tolerance rather than by floating-point variation. It could be widened by producing the reference output with a tighter solve.
\item \textbf{The \nolinkurl{ccd_ee} submission at $0.57$ is also correct.} All but one of its time-of-impact values agree with the reference to within one unit in the last place. The remaining edge crosses its partner about $10^{-6}$ from an endpoint. At that edge, the submission forms the difference between the two edges from absolute coordinates, and the reference forms it from a precomputed relative offset. The resulting cancellation rejects this crossing and returns the edge's next impact instead.
\item \textbf{The \nolinkurl{pic_flip} submission that fails at $7.6\times$ has a real bug.} It updates the level set only in the cells within one cell of each particle, whereas the specified level set can be lowered by a particle at any cell centre within $r+\Delta x \approx 1.87$ cells, which spans two cells on each side. The missed cells are the air cells adjacent to the free surface, which set the free-surface fractions of the pressure system. Widening the three loops to two cells, with no other change, brings the submission to $0.16$, in line with the passing submissions.
\item \textbf{The \nolinkurl{self_collision_xpbd} submission that fails at $2.2\times$ also has a real bug.} Its hashed cell lookup can visit the same bucket twice, which records a few contact pairs twice, contrary to the specification that each pair appears once.
\end{itemize}

Taken together, correct variation in these experiments stays below about $0.6$ of the tolerance. Every planted bug exceeds it, most by orders of magnitude. The narrowest gap, on \nolinkurl{mc_r}, comes from a solver tolerance inherited by the reference, not from floating-point variation.

\clearpage
\section{Case Studies of Generated Implementations}
\label{app:case_studies}

We examine two tasks that illustrate complementary optimization challenges: organizing particle-to-grid accumulation, and reducing iteration costs in a global solve. The cases come from the same evaluation as Table~\ref{tab:main_results} and were evaluated on the NVIDIA GeForce RTX 4090. Table~\ref{tab:case_studies} reports speedup as reference execution time divided by generated execution time on hidden inputs, consistent with the definition used for $\mathrm{fast}_p$; all four submissions pass numerical checks and audits. The discussion combines evaluation results with inspection of final CUDA code and development traces.

\begin{table}[ht]
\centering
\small
\setlength{\tabcolsep}{5pt}
\renewcommand{\arraystretch}{1.12}
\caption{Final performance of the selected implementations on hidden inputs. For passing submissions, higher $t_{\mathrm{ref}}/t_{\mathrm{gen}}$ is better, and $1\times$ matches the reference.}
\label{tab:case_studies}
\begin{tabular*}{\textwidth}{@{\extracolsep{\fill}}llr@{}}
\toprule
\textbf{Task} & \textbf{Model} & $\boldsymbol{t_{\mathrm{ref}}/t_{\mathrm{gen}}}$ $\uparrow$ \\
\midrule
\multirow{2}{*}{PIC/FLIP particle-to-grid transfer}
 & Claude-Opus-5 & $0.556\times$ \\
 & Gemini-3.5-Flash & $0.134\times$ \\
\midrule
\multirow{2}{*}{Dirichlet Poisson solve}
 & Claude-Opus-5 & $0.049\times$ \\
 & Gemini-3.5-Flash & $0.036\times$ \\
\bottomrule
\end{tabular*}
\end{table}

\subsection{PIC/FLIP Particle-to-Grid Transfer: Organizing Accumulation}
\label{app:case_p2g}

The \texttt{p2g\_pic\_flip} task transfers particle velocities onto a staggered MAC grid using trilinear weights. Each particle contributes mass and momentum to eight neighboring nodes for each velocity component, followed by normalization of momentum by mass. Nearby particles update overlapping grid nodes, making the organization of accumulation central to performance.

Gemini sorts particle indices by grid-cell ID using CUB radix sort, then processes particles in that order. However, each thread still gathers particle data indirectly from the original arrays and performs global atomic additions for every particle-node contribution. Sorting improves spatial ordering without aggregating contributions before they reach global memory.

Opus groups particles into $8\times8\times8$ cell tiles through counting and prefix sums, then packs their positions and velocities into contiguous tile buffers. One CUDA block handles each tile, accumulating mass and momentum in shared memory over a $10\times10\times10$ region that includes the halo. The block then merges these values into global accumulators. This reduces global atomic traffic, but each particle still performs atomic updates within the shared-memory tile. All grouping, packing, accumulation, and normalization occur inside the timed \texttt{compute()}.

Opus and Gemini achieve speedups of $0.556\times$ and $0.134\times$ relative to the reference, respectively. The reference further sorts particles by cell within each tile and sums each cell's particle contributions in registers before merging them into shared memory. Its fixed-capacity tile buckets also avoid a global counting-sort pass when capacity is sufficient, with a fallback for overflowing tiles. These differences illustrate why spatial grouping alone is insufficient: the granularity of aggregation and the cost of organizing particles both matter, even after most global atomics have been replaced by shared-memory accumulation.

\subsection{Dirichlet Poisson Solve: Iteration Cost and Solver Choice}
\label{app:case_poisson}

This task solves a variable-coefficient Poisson system with homogeneous Dirichlet boundary conditions on a $256^3$ grid. The operator is symmetric positive definite on the active cells, and the solution must satisfy a relative residual tolerance of $10^{-6}$. Unlike fixed-iteration updates, the total work depends on both the cost of each iteration and the convergence behavior of the chosen solver.

Gemini implements conjugate gradients with diagonal Jacobi preconditioning. Each iteration uses separate kernels for the matrix-vector product, dot products, solution and residual updates, and search-direction update. Two scalar reductions are copied back to the host in a typical iteration to compute the CG coefficients, introducing repeated synchronization. Periodic convergence checks recompute the true residual before accepting the solution and restart CG if necessary.

Opus also uses diagonal preconditioning, applying it through symmetric scaling of the system inside \texttt{compute()}. This folds the preconditioner into an operator with unit diagonal. One main kernel fuses the deferred solution update, search-direction update, matrix-vector product, and dot-product accumulation; another updates the residual and accumulates its norm. Additional small reduction kernels keep the CG coefficients on the device. Host transfers are used for initialization, convergence checks every ten iterations, and true-residual verification, rather than for the coefficients of every iteration.

The final implementations achieve speedups of $0.049\times$ and $0.036\times$ relative to the reference for Opus and Gemini, respectively. Opus reduces memory passes and host synchronization, but both implementations retain a diagonal preconditioner, whereas the expert reference uses multigrid-preconditioned CG. Their large remaining gaps illustrate the limits of optimizing individual iterations without also improving the solver's convergence behavior. Efficient global solves require numerical algorithm selection and GPU execution to be optimized together.

\clearpage
\section{Benchmark Tasks}
\label{app:task_list}

\subsection{Task List}
\label{app:task_list_table}

The benchmark covers fluid dynamics, deformable solids, and granular materials. The fluid tasks span Eulerian grid-based solvers~\citep{zehnder2018reflection}, hybrid particle-grid methods such as particle-in-cell/fluid-implicit-particle (PIC/FLIP) and affine particle-in-cell (APIC)~\citep{jiang2015apic}, particle-based methods such as position-based fluids (PBF)~\citep{miles2013pbf}, and the lattice Boltzmann method (LBM)~\citep{li2026lbm}. They include both local operations, such as advection, particle-grid transfer, lattice collision and streaming, and phase-field evolution, and global or iterative computations, such as pressure projection and implicit viscosity. The solid and granular tasks cover mass-spring systems~\citep{liu2013ms}, the finite element method (FEM)~\citep{sifakis2012fem}, the material point method (MPM)~\citep{Stomakhin2013snow}, extended position-based dynamics (XPBD)~\citep{Miles2016xpbd}, continuous collision detection (CCD)~\citep{Brochu2022ccd}, and the discrete element method (DEM)~\citep{lu2022dem}. Together, these tasks exercise explicit and implicit integration, projective and constraint-based solvers, contact handling, and coupled multi-stage simulation pipelines.

Table~\ref{tab:task_list} lists all 50 benchmark tasks under the six categories used in Section~\ref{sec:category_results}. Test names match the task directories in the benchmark codebase.

\begingroup
\footnotesize
\setlength{\tabcolsep}{4pt}
\renewcommand{\arraystretch}{1.15}
\setlength{\LTleft}{0pt}
\setlength{\LTright}{0pt}
\newlength{\tasklistwidth}
\setlength{\tasklistwidth}{\dimexpr\textwidth-8\tabcolsep-5\arrayrulewidth\relax}
\begin{longtable}{|>{\centering\arraybackslash}p{0.05\tasklistwidth}|>{\raggedright\arraybackslash}p{0.22\tasklistwidth}|>{\raggedright\arraybackslash}p{0.34\tasklistwidth}|>{\raggedright\arraybackslash}p{0.39\tasklistwidth}|}
\caption{Complete GPUPhysBench task list. Descriptions summarize the computation specified in each task prompt.}\label{tab:task_list}\\
\hline
\textbf{No.} & \textbf{Category} & \textbf{Test name} & \textbf{Description} \\ \hline
\endfirsthead
\caption[]{Complete GPUPhysBench task list (continued).}\\
\hline
\textbf{No.} & \textbf{Category} & \textbf{Test name} & \textbf{Description} \\ \hline
\endhead
\endfoot
\endlastfoot
1 & Global Solves and Pressure Projection & \nolinkurl{poisson_dirichlet} & Solve a Poisson system with homogeneous Dirichlet boundaries using preconditioned CG. \\ \hline
2 & Global Solves and Pressure Projection & \nolinkurl{poisson_neumann} & Solve a pure-Neumann Poisson system using preconditioned CG and return a zero-mean solution. \\ \hline
3 & Global Solves and Pressure Projection & \nolinkurl{viscosity_implicit} & Solve implicit viscous diffusion for velocity components on a MAC grid. \\ \hline
4 & Global Solves and Pressure Projection & \nolinkurl{mass_spring_newtonian_implicit} & Advance an implicit mass-spring step using Newton's method. \\ \hline
5 & Global Solves and Pressure Projection & \nolinkurl{fem_newtonian_implicit} & Advance an implicit tetrahedral FEM step using Newton's method. \\ \hline
6 & Global Solves and Pressure Projection & \nolinkurl{mass_spring_pd} & Advance a mass-spring step using projective dynamics local-global iterations. \\ \hline
7 & Global Solves and Pressure Projection & \nolinkurl{fem_pd} & Advance a tetrahedral FEM step using projective dynamics local-global iterations. \\ \hline
8 & Global Solves and Pressure Projection & \nolinkurl{stable_fluids} & Advance a MAC-grid fluid step with advection, implicit viscosity, and pressure projection. \\ \hline
9 & Global Solves and Pressure Projection & \nolinkurl{mc_r} & Advance a MAC-grid fluid step using MacCormack advection, reflection, and two pressure projections. \\ \hline
10 & Position Constraint Solving & \nolinkurl{solve_constraint_pbf} & Iteratively correct particle positions to satisfy PBF density constraints. \\ \hline
11 & Position Constraint Solving & \nolinkurl{pbf} & Advance a full PBF step with prediction, density-constraint solving, velocity update, vorticity confinement, and XSPH viscosity. \\ \hline
12 & Position Constraint Solving & \nolinkurl{solve_constraint_xpbd} & Solve XPBD spring constraints by iteratively correcting particle positions. \\ \hline
13 & Position Constraint Solving & \nolinkurl{self_collision_xpbd} & Resolve cloth particle self-collisions with XPBD contact constraints. \\ \hline
14 & Position Constraint Solving & \nolinkurl{xpbd} & Advance a cloth mass-spring timestep using XPBD with spring and self-collision contact constraints. \\ \hline
15 & Local Grid and Lattice Computations & \nolinkurl{advect_rk1} & Advect grid velocities using semi-Lagrangian transport with Euler backtracing. \\ \hline
16 & Local Grid and Lattice Computations & \nolinkurl{advect_rk2} & Advect grid velocities using semi-Lagrangian transport with midpoint RK2 backtracing. \\ \hline
17 & Local Grid and Lattice Computations & \nolinkurl{advect_rk3} & Advect grid velocities using semi-Lagrangian transport with Ralston RK3 backtracing. \\ \hline
18 & Local Grid and Lattice Computations & \nolinkurl{advect_rk4} & Advect grid velocities using semi-Lagrangian transport with RK4 backtracing. \\ \hline
19 & Local Grid and Lattice Computations & \nolinkurl{advect_mc} & Advect grid velocities using MacCormack correction and RK2 backtracing. \\ \hline
20 & Local Grid and Lattice Computations & \nolinkurl{stream_d3q19} & Stream D3Q19 lattice distributions to neighboring grid cells. \\ \hline
21 & Local Grid and Lattice Computations & \nolinkurl{stream_d3q27} & Stream D3Q27 lattice distributions to neighboring grid cells. \\ \hline
22 & Local Grid and Lattice Computations & \nolinkurl{trt_collision_d3q19} & Apply two-relaxation-time collision to D3Q19 lattice distributions. \\ \hline
23 & Local Grid and Lattice Computations & \nolinkurl{trt_collision_d3q27} & Apply two-relaxation-time collision to D3Q27 lattice distributions. \\ \hline
24 & Local Grid and Lattice Computations & \nolinkurl{lbm_d3q19} & Advance a D3Q19 lattice Boltzmann step with TRT collision and streaming. \\ \hline
25 & Local Grid and Lattice Computations & \nolinkurl{lbm_d3q27} & Advance a D3Q27 lattice Boltzmann step with TRT collision and streaming. \\ \hline
26 & Local Grid and Lattice Computations & \nolinkurl{cahn_hilliard} & Advance the Cahn--Hilliard phase field using a lattice Boltzmann step. \\ \hline
27 & Local Grid and Lattice Computations & \nolinkurl{surface_tension_sdf} & Compute surface-tension forces from a level-set signed distance field. \\ \hline
28 & Local Grid and Lattice Computations & \nolinkurl{surface_tension_phase_field} & Compute surface-tension forces from the phase field and its chemical potential. \\ \hline
29 & Particle-Grid Methods & \nolinkurl{p2g_pic_flip} & Transfer particle velocities to a MAC grid for PIC/FLIP simulation. \\ \hline
30 & Particle-Grid Methods & \nolinkurl{g2p_pic_flip} & Update particle velocities by blending PIC and FLIP grid-to-particle transfers. \\ \hline
31 & Particle-Grid Methods & \nolinkurl{p2g_apic} & Transfer particle velocities and affine momentum to the APIC grid. \\ \hline
32 & Particle-Grid Methods & \nolinkurl{g2p_apic} & Gather grid velocities and the per-particle affine state $B$ for APIC. \\ \hline
33 & Particle-Grid Methods & \nolinkurl{p2g_mpm} & Scatter particle mass, momentum, and stress forces to the MPM grid. \\ \hline
34 & Particle-Grid Methods & \nolinkurl{g2p_mpm} & Update MPM particle velocity, deformation gradient, and position from grid fields. \\ \hline
35 & Particle-Grid Methods & \nolinkurl{velocity_gradient_mpm} & Evaluate the velocity gradient at MPM particles from grid velocities. \\ \hline
36 & Particle-Grid Methods & \nolinkurl{apic} & Advance a full APIC fluid step with transfers, gravity, free-surface pressure projection, and particle advection. \\ \hline
37 & Particle-Grid Methods & \nolinkurl{pic_flip} & Advance a full PIC/FLIP fluid step with transfers, gravity, free-surface pressure projection, and particle advection. \\ \hline
38 & Particle-Grid Methods & \nolinkurl{mpm_explicit} & Advance an explicit MPM step through particle-grid transfers and grid dynamics. \\ \hline
39 & Particle-Grid Methods & \nolinkurl{mpm_semi_implicit} & Advance a semi-implicit MPM step with an implicit grid-velocity solve. \\ \hline
40 & Geometric Queries and Collision Detection & \nolinkurl{ccd_vv} & Detect continuous vertex-vertex collisions during cloth motion. \\ \hline
41 & Geometric Queries and Collision Detection & \nolinkurl{ccd_ve} & Detect continuous vertex-edge collisions during cloth motion. \\ \hline
42 & Geometric Queries and Collision Detection & \nolinkurl{ccd_vf} & Detect continuous vertex-face collisions during cloth motion. \\ \hline
43 & Geometric Queries and Collision Detection & \nolinkurl{ccd_ee} & Detect continuous edge-edge collisions during cloth motion. \\ \hline
44 & Geometric Queries and Collision Detection & \nolinkurl{particle_sdf} & Build a clamped grid signed distance field from particle-centered spheres. \\ \hline
45 & Local Interaction Updates & \nolinkurl{mass_spring_explicit} & Compute spring forces and explicitly advance particle positions and velocities. \\ \hline
46 & Local Interaction Updates & \nolinkurl{fem_explicit} & Compute tetrahedral elastic forces and explicitly advance mesh vertices. \\ \hline
47 & Local Interaction Updates & \nolinkurl{contact_dem} & Evaluate normal contact forces between overlapping spherical DEM particles. \\ \hline
48 & Local Interaction Updates & \nolinkurl{dem} & Advance a DEM timestep with frictional particle contacts and state integration. \\ \hline
49 & Local Interaction Updates & \nolinkurl{viscosity_pbf} & Apply XSPH viscosity to PBF particle velocities using neighboring particles. \\ \hline
50 & Local Interaction Updates & \nolinkurl{vorticity_confinement_pbf} & Compute vorticity confinement and update PBF particle velocities. \\ \hline
\end{longtable}
\endgroup

\subsection{Task Inputs and Correctness Checks}
\label{app:task_inputs}

Table~\ref{tab:task_inputs} gives, for every task, the problem size, how the hidden inputs differ from the public ones, and each correctness check with its absolute tolerance. The hidden inputs keep the problem size of the public inputs, except for small changes in particle or unknown counts on five tasks, so that timings on the two sets are comparable. They differ in their data. Every task except \nolinkurl{surface_tension_sdf} draws a different random seed or initial condition, and that task instead moves its ellipsoid. Sixteen tasks change the geometry, such as obstacles, cut cells, drop or cloth layouts, and collision sheets, and four change a coefficient that the solver receives. The table is generated from the benchmark's input generators and evaluators by \texttt{analysis/task\_inventory/input\_spec.py}.

\begingroup
\footnotesize
\setlength{\tabcolsep}{4pt}
\renewcommand{\arraystretch}{1.15}
\setlength{\LTleft}{0pt}
\setlength{\LTright}{0pt}
\newlength{\taskinputswidth}
\setlength{\taskinputswidth}{\dimexpr\textwidth-8\tabcolsep-5\arrayrulewidth\relax}
\begin{longtable}{|>{\raggedright\arraybackslash}p{0.17\taskinputswidth}|>{\raggedright\arraybackslash}p{0.25\taskinputswidth}|>{\raggedright\arraybackslash}p{0.29\taskinputswidth}|>{\raggedright\arraybackslash}p{0.29\taskinputswidth}|}
\caption{Inputs, hidden-input variation and correctness checks of every task. Each call advances or evaluates one step. Sizes are those the driver passes to the submission; the hidden inputs have the same size unless a hidden size is given. The hidden inputs come from the public generator with the listed constants changed (public$\to$hidden); grids, particle lattices, mesh topology, material parameters, time steps and solver settings are otherwise kept. A check passes when its value is below the listed absolute tolerance on both input sets. Unless stated otherwise it is the FP64 relative error $\|g-r\|_2/\|r\|_2$ of the submission output $g$ against the reference output $r$ on the same inputs; $\Delta q=q-q_0$ is the change from the input, and ($\times k$) marks $k$ separately checked components.}\label{tab:task_inputs}\\
\hline
\textbf{Task} & \textbf{Public / hidden size} & \textbf{Hidden-input variation} & \textbf{Checks and tolerances} \\ \hline
\endfirsthead
\caption[]{Task inputs and correctness checks (continued).}\\
\hline
\textbf{Task} & \textbf{Public / hidden size} & \textbf{Hidden-input variation} & \textbf{Checks and tolerances} \\ \hline
\endhead
\endfoot
\endlastfoot
\multicolumn{4}{|l|}{\emph{Local Grid and Lattice Computations}} \\ \hline
\nolinkurl{advect_rk1} & MAC velocity grid $256^3$ & seed; Fourier modes 16$\to$20 (same RMS, same $k_{\max}$) & rel.\ $L_2$ of $u_x,u_y,u_z$ ($\times3$): $10^{-6}$ \\ \hline
\nolinkurl{advect_rk2} & MAC velocity grid $256^3$ & seed; Fourier modes 16$\to$20 (same RMS, same $k_{\max}$) & rel.\ $L_2$ of $u_x,u_y,u_z$ ($\times3$): $10^{-6}$ \\ \hline
\nolinkurl{advect_rk3} & MAC velocity grid $256^3$ & seed; Fourier modes 16$\to$20 (same RMS, same $k_{\max}$) & rel.\ $L_2$ of $u_x,u_y,u_z$ ($\times3$): $10^{-6}$ \\ \hline
\nolinkurl{advect_rk4} & MAC velocity grid $256^3$ & seed; Fourier modes 16$\to$20 (same RMS, same $k_{\max}$) & rel.\ $L_2$ of $u_x,u_y,u_z$ ($\times3$): $10^{-6}$ \\ \hline
\nolinkurl{advect_mc} & MAC velocity grid $256^3$ & seed; Fourier modes 16$\to$20 (same RMS, same $k_{\max}$) & rel.\ $L_2$ of $u_x,u_y,u_z$ ($\times3$): $10^{-4}$ \\ \hline
\nolinkurl{stream_d3q19} & D3Q19 lattice $288\times256\times224$ & seed; ABC flow 4$\to$3 periods, Mach 0.05$\to$0.043; density amplitude 0.05$\to$0.06; non-eq.\ perturbation 0.3$\to$0.36 & rel.\ $L_2$ of $f$: $10^{-9}$ \\ \hline
\nolinkurl{stream_d3q27} & D3Q27 lattice $288\times256\times224$ & seed; ABC flow 4$\to$3 periods, Mach 0.05$\to$0.043; density amplitude 0.05$\to$0.06; non-eq.\ perturbation 0.3$\to$0.36 & rel.\ $L_2$ of $f$: $10^{-9}$ \\ \hline
\nolinkurl{trt_collision_d3q19} & D3Q19 lattice $256^3$ & seed; ABC flow 4$\to$3 periods, Mach 0.05$\to$0.043; density amplitude 0.05$\to$0.06; non-eq.\ part 0.3$\to$0.26 & rel.\ $L_2$ of $f$: $5\times10^{-6}$ \\ \hline
\nolinkurl{trt_collision_d3q27} & D3Q27 lattice $256^3$ & seed; ABC flow 4$\to$3 periods, Mach 0.05$\to$0.043; density amplitude 0.05$\to$0.06; non-eq.\ part 0.2$\to$0.17 & rel.\ $L_2$ of $f$: $5\times10^{-6}$ \\ \hline
\nolinkurl{lbm_d3q19} & D3Q19 lattice $288\times256\times224$ & seed; ABC flow 4$\to$3 periods, Mach 0.05$\to$0.043; density amplitude 0.05$\to$0.06; non-eq.\ part 0.25$\to$0.21 & rel.\ $L_2$ of $f$: $5\times10^{-6}$ \\ \hline
\nolinkurl{lbm_d3q27} & D3Q27 lattice $288\times256\times224$ & seed; ABC flow 4$\to$3 periods, Mach 0.05$\to$0.043; density amplitude 0.05$\to$0.06; non-eq.\ part 0.18$\to$0.15 & rel.\ $L_2$ of $f$: $5\times10^{-6}$ \\ \hline
\nolinkurl{cahn_hilliard} & D3Q19 lattice $272\times256\times240$ + velocity field & seed (new drop layout); drop radii [14,36]$\to$[12,32] cells; ABC flow 4$\to$3 periods, Mach 0.05$\to$0.042; non-eq.\ part 0.3$\to$0.26 & rel.\ $L_2$ of $h$ over all directions: $3\times10^{-5}$ \\ \hline
\nolinkurl{surface_tension_sdf} & level set $512^3$ & no seed; ellipsoid centre, semi-axes (aspect 1.6$\to$1.48) and tilt all moved & rel.\ $L_2$ of force: $10^{-4}$ \\ \hline
\nolinkurl{surface_tension_phase_field} & phase field $544\times512\times480$ & seed (new drop layout); drop radii [16,44]$\to$[14,38] cells & rel.\ $L_2$ of force: $5\times10^{-4}$ \\ \hline
\multicolumn{4}{|l|}{\emph{Particle--Grid Methods}} \\ \hline
\nolinkurl{p2g_pic_flip} & 33,021,538 particles $\to$ MAC grid $256^3$ \newline \emph{Hidden:} 33,017,033 particles & seed (jitter, order, velocities); jitter margin 0.01$\to$0.015 cell & rel.\ $L_2$ of grid $u_x,u_y,u_z$ ($\times3$): $10^{-6}$ \\ \hline
\nolinkurl{g2p_pic_flip} & old/new MAC grids $256^3$ $\to$ 33,021,538 particles \newline \emph{Hidden:} 33,017,033 particles & particle and grid seeds; jitter margin 0.01$\to$0.015 cell & rel.\ $L_2$ of particle $u$ ($\times3$): $10^{-6}$ \\ \hline
\nolinkurl{p2g_apic} & 33,021,538 particles $\to$ MAC grid $256^3$ \newline \emph{Hidden:} 33,017,033 particles & seed; affine-matrix scale 1.0$\to$0.8; jitter margin 0.01$\to$0.015 cell & rel.\ $L_2$ of grid $u_x,u_y,u_z$ ($\times3$): $2\times10^{-6}$ \\ \hline
\nolinkurl{g2p_apic} & MAC grid $256^3$ $\to$ 33,021,538 particles \newline \emph{Hidden:} 33,017,033 particles & particle and grid seeds; jitter margin 0.01$\to$0.015 cell & rel.\ $L_2$ of $u$ ($\times3$) and of each affine entry $B$ ($\times9$): $10^{-6}$ \\ \hline
\nolinkurl{p2g_mpm} & 16.86M particles $\to$ grid $256^3$ & seed (velocities, $F$ jitter, order); $F$ jitter 0.05$\to$0.06 & rel.\ $L_2$ of grid $u_x,u_y,u_z$ ($\times3$): $10^{-5}$ \\ \hline
\nolinkurl{g2p_mpm} & grid $256^3$ $\to$ 16.86M particles & seed ($F$ jitter, grid velocity, order); $F$ jitter 0.05$\to$0.06 & rel.\ $L_2$ of $u$: $10^{-5}$; of $F$: $10^{-5}$; of $\Delta x$: $10^{-3}$ \\ \hline
\nolinkurl{velocity_gradient_mpm} & grid $256^3$ $\to$ 16.78M particles & seeds only (particle order, grid velocity) & rel.\ $L_2$ of each $\nabla v$ entry ($\times9$): $10^{-4}$ \\ \hline
\nolinkurl{apic} & 33.29M particles, MAC grid $256^3$; pressure-solve tol $10^{-5}$ & seed; swirl z-fade 0.5$\to$0.4 and z-part 0.3$\to$0.38; affine scale 1.0$\to$0.8; jitter margin 0.01$\to$0.015 & rel.\ $L_2$ of $\Delta u$ ($\times3$), of $x-(x_0+\Delta t\,u_0)$ ($\times3$) and of each $C$ entry ($\times9$): $3\times10^{-3}$ \\ \hline
\nolinkurl{pic_flip} & 33.29M particles, MAC grid $256^3$; pressure-solve tol $10^{-4}$ & seed; swirl z-fade 0.5$\to$0.4 and z-part 0.3$\to$0.38; particle noise 0.15$\to$0.18; jitter margin 0.01$\to$0.015 & rel.\ $L_2$ of $\Delta u$ and $\Delta x$ ($\times6$): $8\times10^{-3}$ \\ \hline
\nolinkurl{mpm_explicit} & 16.86M particles, grid $256^3$ & seed (velocities, $F$ jitter, order); $F$ jitter 0.05$\to$0.06 & rel.\ $L_2$ of $u$: $10^{-5}$; of $F$: $10^{-5}$; of $\Delta x$: $10^{-3}$ \\ \hline
\nolinkurl{mpm_semi_implicit} & 16.86M particles, grid $256^3$; solver tol $10^{-5}$ & seed (velocities, $F$ jitter, order); $F$ jitter 0.05$\to$0.06 & implicit grid-momentum residual of the returned grid velocity, recomputed in FP64 ($2\times$ solver tol): $2\times10^{-5}$; max rel.\ $L_2$ mismatch of $u,F,\Delta x$ vs.\ an FP64 G2P from that grid: $10^{-4}$ \\ \hline
\multicolumn{4}{|l|}{\emph{Local Interaction Updates}} \\ \hline
\nolinkurl{mass_spring_explicit} & 889K particles ($96^3$ lattice + 4096 free), 7.82M springs & seed; deformation/velocity field 4$\to$5 periods; strain 0.01$\to$0.012; jitter 0.01$\to$0.008; velocity noise 0.1$\to$0.12 & rel.\ $L_2$ of $x$: $10^{-6}$; of $v$: $5\times10^{-5}$ \\ \hline
\nolinkurl{fem_explicit} & 913K vertices, 5.31M tets ($96^3$ cells) & seed; deformation/velocity field 2$\to$3 periods; strain 0.08$\to$0.07; jitter 0.05$\to$0.04; velocity noise 0.1$\to$0.12 & rel.\ $L_2$ of $x$: $10^{-5}$; of $v$: $10^{-4}$ \\ \hline
\nolinkurl{contact_dem} & 2.06M grains ($144\times128\times112$ lattice) & seed; jitter 0.12$\to$0.10; radii [0.50,0.60]$\to$[0.485,0.605]; shear 0.01$\to$0.013; velocity noise 0.5$\to$0.48 & rel.\ $L_2$ of contact force: $4\times10^{-4}$ \\ \hline
\nolinkurl{dem} & 2.06M grains ($144\times128\times112$ lattice) & seed; jitter 0.12$\to$0.10; radii [0.50,0.60]$\to$[0.485,0.605]; shear 0.01$\to$0.013; velocity noise 0.5$\to$0.45; spin 0.6$\to$0.72 & rel.\ $L_2$ of $\Delta x$: $3\times10^{-4}$; of $\Delta v$: $2\times10^{-4}$; of $\Delta\omega$: $4\times10^{-4}$ \\ \hline
\nolinkurl{viscosity_pbf} & 4.10M particles ($160^3$ lattice) & seed; jitter 0.15$\to$0.12; ABC wavelength 8$\to$7 spacings; noise 0.15$\to$0.18; block corner 10$\to$12.5 & rel.\ $L_2$ of $\Delta v$: $10^{-4}$ \\ \hline
\nolinkurl{vorticity_confinement_pbf} & 4.10M particles ($160^3$ lattice) & seed; jitter 0.15$\to$0.13; ABC wavelength 16$\to$13 spacings; block corner 10$\to$12.5; confinement strength re-derived (0.315$\to$0.256) & rel.\ $L_2$ of $\Delta v$: $10^{-3}$ \\ \hline
\multicolumn{4}{|l|}{\emph{Position Constraint Solving}} \\ \hline
\nolinkurl{solve_constraint_pbf} & 4.10M particles ($160^3$ lattice); 5 iterations & seed; density wave amplitude 0.3$\to$0.26, wavelength 16$\to$20 spacings; jitter 0.15$\to$0.12 & rel.\ $L_2$ of $\Delta x$: $10^{-3}$ \\ \hline
\nolinkurl{pbf} & 4.10M particles ($160^3$ lattice); 5 iterations & seed; density wave 0.15$\to$0.18, 16$\to$20 spacings; ABC wavelength 8$\to$10 (vorticity strength 0.083$\to$0.104); jitter 0.15$\to$0.12 & rel.\ $L_2$ of $\Delta x$: $2\times10^{-3}$; of $v$: $3\times10^{-4}$ \\ \hline
\nolinkurl{solve_constraint_xpbd} & 1.05M particles ($1024^2$ sheet + 4096 free), 6.28M springs; 8 iterations & seed; roll start radius 5$\to$6, turn gap 0.45$\to$0.42; strain 0.005$\to$0.006; 4$\to$5 periods; jitter 0.005$\to$0.004; noise 0.1$\to$0.12 & rel.\ $L_2$ of $\Delta x$: $4\times10^{-3}$; spring-free particles unmoved: $10^{-6}$ \\ \hline
\nolinkurl{self_collision_xpbd} & 1.05M particles ($1024^2$ sheet, 8 folds); 8 iterations & seed; layer gap 0.70$\to$0.68; warp 0.35$\to$0.40 over 3$\to$2 periods (new contact patches); jitter 0.02$\to$0.018 & rel.\ $L_2$ of $\Delta x$: $2\times10^{-3}$; particles the reference leaves in place unmoved: $10^{-6}$ \\ \hline
\nolinkurl{xpbd} & 1.05M particles ($1024^2$ sheet + 4096 free), 6.28M springs; 8 iterations & seed; roll start radius 5$\to$6, turn gap 0.45$\to$0.42; strain 0.005$\to$0.006; 4$\to$5 periods; jitter 0.005$\to$0.004; noise 0.1$\to$0.12 & rel.\ $L_2$ of $x$: $2.3\times10^{-7}$; of $v$: $2.6\times10^{-4}$; free particles vs.\ $x_0+\Delta t\,v_0$: $10^{-5}$ \\ \hline
\multicolumn{4}{|l|}{\emph{Geometric Queries and Collision Detection}} \\ \hline
\nolinkurl{ccd_vv} & 7.84M vertices (4 sheets of $1400^2$) & seed; sheet offset, waviness, wavenumber, phase step and drive period (a different region interpenetrates); jitter 0.15$\to$0.16 & rel.\ $L_2$ of $\mathrm{toi}-1$: $3\times10^{-2}$ \\ \hline
\nolinkurl{ccd_ve} & 262K vertices, 782K edges (4 sheets of $256^2$) & seed; sheet offset, waviness, wavenumber, phase step and drive period (a different region interpenetrates) & rel.\ $L_2$ of $\mathrm{toi}-1$: $10^{-3}$ \\ \hline
\nolinkurl{ccd_vf} & 590K vertices, 1.17M triangles (4 sheets of $384^2$) & seed; sheet offset, waviness, wavenumber, phase step and drive period (a different region interpenetrates); jitter 0.06$\to$0.05 & rel.\ $L_2$ of $\mathrm{toi}-1$: $10^{-4}$ \\ \hline
\nolinkurl{ccd_ee} & 410K vertices, 1.22M edges (4 sheets of $320^2$) & seed; sheet offset, waviness, wavenumber, phase step and drive period (a different region interpenetrates); jitter 0.06$\to$0.07 & rel.\ $L_2$ of $\mathrm{toi}-1$: $4\times10^{-6}$ \\ \hline
\nolinkurl{particle_sdf} & 8.79M particles $\to$ SDF grid $256^3$ & seed; domain length 1.0$\to$0.8 with ball radius 0.25$\to$0.2 (cell size 1/256$\to$1/320, same count) & max abs.\ error / cell size: $10^{-4}$ \\ \hline
\multicolumn{4}{|l|}{\emph{Global Solves and Pressure Projection}} \\ \hline
\nolinkurl{poisson_dirichlet} & grid $256^3$, 8.39M unknowns; PCG to rel.\ residual $10^{-6}$ & seed (face conductances, manufactured solution); conductance range [0.1,1]$\to$[0.12,1.15] & true residual $\|Ax-b\|/\|b\|$ (= solver tol): $10^{-6}$; $x=0$ exactly outside the domain \\ \hline
\nolinkurl{poisson_neumann} & grid $256^3$, 15,118,336 unknowns; PCG to rel.\ residual $10^{-6}$ \newline \emph{Hidden:} 14,505,984 unknowns & seed; cylinder radius 0.18$\to$0.21 of the grid (new domain, cut-cell coefficients) & true residual $\|Ax-b\|/\|b\|$ (= solver tol): $10^{-6}$; $|\mathrm{mean}(x)|/\mathrm{rms}(x)$: $10^{-4}$; $x=0$ exactly outside the domain \\ \hline
\nolinkurl{viscosity_implicit} & MAC grid $256^3$ with cut-cell cylinder; solver tol $3\times10^{-4}$ & seed; Taylor--Green modes 48$\to$56; cylinder radius 0.18$\to$0.21 of the grid (new cut cells) & rel.\ $L_2$ of $u_x,u_y,u_z$ ($\times3$): $2.5\times10^{-3}$ \\ \hline
\nolinkurl{mass_spring_newtonian_implicit} & 889K particles ($96^3$ lattice + 4096 free), 7.82M springs; Newton to rel.\ residual $10^{-3}$ & seed; deformation/velocity field 4$\to$3 periods; jitter 0.005$\to$0.006; velocity noise 0.1$\to$0.12 & true FP64 residual $\|g(x)\|/\|g(y)\|$ (tol + $5\times10^{-5}$): $1.05\times10^{-3}$; rel.\ $L_2$ of $v$ vs.\ $(x-x_0)/\Delta t$: $10^{-5}$ \\ \hline
\nolinkurl{fem_newtonian_implicit} & 275K vertices, 1.57M tets ($64^3$ cells); Newton to rel.\ residual $10^{-3}$ & seed; deformation/velocity field 2$\to$3 periods; strain 0.08$\to$0.07; jitter 0.05$\to$0.04; velocity noise 0.1$\to$0.12 & true FP64 residual $\|g(x)\|/\|g(y)\|$ (tol + $10^{-5}$): $1.01\times10^{-3}$; rel.\ $L_2$ of $v$ vs.\ $(x-x_0)/\Delta t$: $6\times10^{-6}$ \\ \hline
\nolinkurl{mass_spring_pd} & 889K particles ($96^3$ lattice + 4096 free), 7.82M springs; 30 iterations & seed; deformation/velocity field 4$\to$3 periods; jitter 0.005$\to$0.006; velocity noise 0.1$\to$0.12 & rel.\ $L_2$ of $x$: $2.5\times10^{-6}$; of $v$: $7\times10^{-4}$; free particles vs.\ $x_0+\Delta t\,v_0$: $10^{-5}$ \\ \hline
\nolinkurl{fem_pd} & 275K vertices, 1.57M tets ($64^3$ cells); 8 iterations & seed; deformation/velocity field 2$\to$3 periods; strain 0.08$\to$0.07; jitter 0.05$\to$0.04; velocity noise 0.1$\to$0.12 & rel.\ $L_2$ of $x$: $6\times10^{-6}$; of $v$: $3\times10^{-4}$ \\ \hline
\nolinkurl{stable_fluids} & MAC grid $256^3$ with cut-cell cylinder; solver tol $10^{-4}$ & seed; Taylor--Green modes 48$\to$56; cylinder radius 0.18$\to$0.21 of the grid (new cut cells) & rel.\ $L_2$ of $\Delta u_x,\Delta u_y,\Delta u_z$ ($\times3$): $7\times10^{-4}$ \\ \hline
\nolinkurl{mc_r} & MAC grid $256^3$ with cut-cell cylinder; 2 projections, solver tol $10^{-4}$ & seed; Taylor--Green modes 48$\to$56; cylinder radius 0.18$\to$0.21 of the grid (new cut cells) & rel.\ $L_2$ of $\Delta u_x,\Delta u_y,\Delta u_z$ ($\times3$): $10^{-3}$ \\ \hline
\end{longtable}
\endgroup

\clearpage
\section{Comparison with GPU Libraries}
\label{app:library_comparison}

We compare reference implementations with established GPU libraries on each task's public input on an NVIDIA GeForce RTX 4090. All times are GPU times, reported as the best of five runs after three warm-up runs. References are timed like the task driver: a fresh solver object per call, with the same warm-up and best-of-five rule.

\subsection{Poisson Solve with Dirichlet Boundaries versus AMGX}
\label{app:library_poisson}

We compare the reference for \nolinkurl{poisson_dirichlet} with the two fastest of three solver configurations shipped with NVIDIA AMGX~2.5.0, a GPU algebraic multigrid (AMG) library. The input is a variable-coefficient 7-point Poisson system with 8.4 million unknowns, solved in FP32 from a zero initial guess to a relative residual of $10^{-6}$. For AMGX, the timed region also includes assembling and uploading the CSR matrix and scattering the solution back to the grid, which together take about 5\,ms. All solvers reach the tolerance.

\begin{table}[ht]
\centering
\small
\setlength{\tabcolsep}{5pt}
\renewcommand{\arraystretch}{1.12}
\caption{Reference solver of \nolinkurl{poisson_dirichlet} versus AMGX. \emph{Total} also includes data preparation (layout conversion, or CSR assembly and upload) and output.}
\label{tab:library_poisson}
\begin{tabular*}{\textwidth}{@{\extracolsep{\fill}}lrrrr@{}}
\toprule
\textbf{Solver} & \textbf{Total (ms)} & \textbf{Setup (ms)} & \textbf{Solve (ms)} & \textbf{Iterations} \\
\midrule
Reference (geometric multigrid PCG) & \textbf{19.1} & 0.3 & 18.0 & 10 \\
AMGX PCG + classical AMG & 171.8 & 101.1 & 65.8 & 17 \\
AMGX PCG + aggregation AMG & 323.5 & 64.8 & 253.6 & 28 \\
\bottomrule
\end{tabular*}
\end{table}

As Table~\ref{tab:library_poisson} shows, the reference is $9.0\times$ faster than the best AMGX configuration. The gap comes from exploiting the structured grid. First, the reference coarsens geometrically, merging fixed $2\times2\times2$ blocks, so every coarse level remains a 7-point stencil and is built in one averaging pass. AMGX instead builds its hierarchy algebraically from the matrix graph. Second, the reference's V-cycle uses red-black Gauss--Seidel smoothing within tiles and repeated coarse-grid corrections, which reduces the iteration count. Third, each iteration is cheaper, because stencil storage reads about $3.5\times$ less matrix data than CSR and the kernels are fused on shared-memory tiles. AMGX targets arbitrary sparse matrices and cannot use this structure, and its configurations were not tuned. The comparison therefore measures the value of specializing to the task rather than a shortcoming of AMGX.

\subsection{Simulation Steps versus Warp and Taichi}
\label{app:library_warp_taichi}

Warp~1.17.0 and Taichi~1.7.2 do not ship the tasks' exact steps. For each comparison, we therefore start from one of the library's own examples, keep its implementation style, and change only the physics to the task's specification. The examples are \texttt{mpm3d.py} for \nolinkurl{mpm_explicit}, \texttt{example\_dem.py} for \nolinkurl{dem}, and the explicit mode of \texttt{implicit\_fem.py} for \nolinkurl{fem_explicit}. Every port matches the reference output to a relative difference below $5\times10^{-7}$. Library times exclude JIT compilation and are measured with CUDA events for Warp and with the kernel profiler for Taichi.

\begin{table}[ht]
\centering
\small
\setlength{\tabcolsep}{5pt}
\renewcommand{\arraystretch}{1.12}
\caption{Reference implementations versus ports of Warp and Taichi examples. For \nolinkurl{fem_explicit}, the step includes computing the rest-state quantities, which the task requires and the Taichi example precomputes once; without them, the Taichi step takes 2.01\,ms.}
\label{tab:library_warp_taichi}
\begin{tabular*}{\textwidth}{@{\extracolsep{\fill}}llrr@{}}
\toprule
\textbf{Task} & \textbf{Library (example)} & \textbf{Reference (ms)} & \textbf{Library (ms)} \\
\midrule
\nolinkurl{mpm_explicit} & Taichi (\texttt{mpm3d.py}) & \textbf{9.08} & 63.12 \\
\nolinkurl{dem} & Warp (\texttt{example\_dem.py}) & \textbf{1.15} & 3.03 \\
\nolinkurl{fem_explicit} & Taichi (\texttt{implicit\_fem.py}) & \textbf{0.95} & 3.59 \\
\bottomrule
\end{tabular*}
\end{table}

As Table~\ref{tab:library_warp_taichi} shows, the references are $2.6$--$7.0\times$ faster. In each case, the difference lies in how memory accesses and accumulation are organized, not in the arithmetic. The library ports process particles or elements in input order and scatter every contribution with global atomics, or read neighbor data indirectly from unsorted arrays. The references first sort the work spatially. They counting-sort particles into grid tiles for MPM and grains into cells for DEM, and they bin tetrahedra into Morton-ordered spatial bins for FEM. They then accumulate each tile or bin in shared memory before a single global update, and pack each grain's state contiguously in cell order so that neighbor loops read cache-friendly records. The MPM reference also allocates grid storage only over the particles' bounding box, and the FEM reference computes the rest-state quantities inside the force kernel instead of storing them.

\clearpage
\section{Simulations Built on Reference Implementations}
\label{app:simulation_visualization}

The reference implementations in GPUPhysBench are complete simulation steps rather than isolated kernels, so they can be advanced over many timesteps to produce full physical simulations. Figure~\ref{fig:simulation_visualization} shows six such simulations, one per row, each driven by the reference implementation of a single benchmark task and run on an NVIDIA GeForce RTX 4090. Each row shows four representative frames, with the simulated time given below each frame. Scene-specific ingredients that lie outside a task's one-step specification, such as gravity, boundaries, contact with scene objects, and material plasticity, are supplied by the scene setup around the reference step.

\paragraph{(a) Vortex-ring collision (\nolinkurl{mc_r}).} Two coaxial vortex rings with slightly perturbed cores collide head-on in a closed box ($128\times256\times256$ MAC grid). They expand radially along the mid-plane until the perturbation grows and breaks them into small-scale vortices. Frames show a volume rendering of the vorticity magnitude.

\paragraph{(b) Water drop (\nolinkurl{pic_flip}).} A drop falls into a tank of still water (25.2 million particles, rendered as spheres colored by speed). The impact forms a crown splash and a cavity, whose collapse drives a Worthington jet.

\paragraph{(c) K\'arm\'an vortex street (\nolinkurl{stable_fluids}).} Flow past a cylinder at a Reynolds number of 1000 ($512\times256\times128$ grid) sheds two rows of alternating vortices. Frames show the spanwise vorticity $\omega_z$ on the mid-span plane (orange and teal for opposite signs), with the flow from left to right. The shedding Strouhal number is $0.197$, close to the experimental value of about $0.2$.

\paragraph{(d) Jelly cube (\nolinkurl{fem_explicit}).} A soft Neo-Hookean cube (48,000 tetrahedra) is dropped onto the ground. It squashes on impact, bounces while wobbling, and recovers its shape.

\paragraph{(e) Snowball (\nolinkurl{mpm_explicit}).} A snowball (1.1 million material points) hits the ground obliquely. The contact patch compacts while the rest shatters and sprays forward. The snow is rendered volumetrically.

\paragraph{(f) Cloth on a sphere (\nolinkurl{xpbd}).} A $160\times160$-particle mass-spring cloth with self-contact falls onto a fixed sphere and drapes over it in folds.

\newcommand{\simlabels}[4]{%
  {\scriptsize\makebox[0.244\linewidth]{#1}\hfill\makebox[0.244\linewidth]{#2}\hfill
   \makebox[0.244\linewidth]{#3}\hfill\makebox[0.244\linewidth]{#4}}}

\begin{figure}[p]
\centering
\small
\textbf{(a)} Vortex-ring collision (\nolinkurl{mc_r})\\[2pt]
\includegraphics[width=0.244\linewidth]{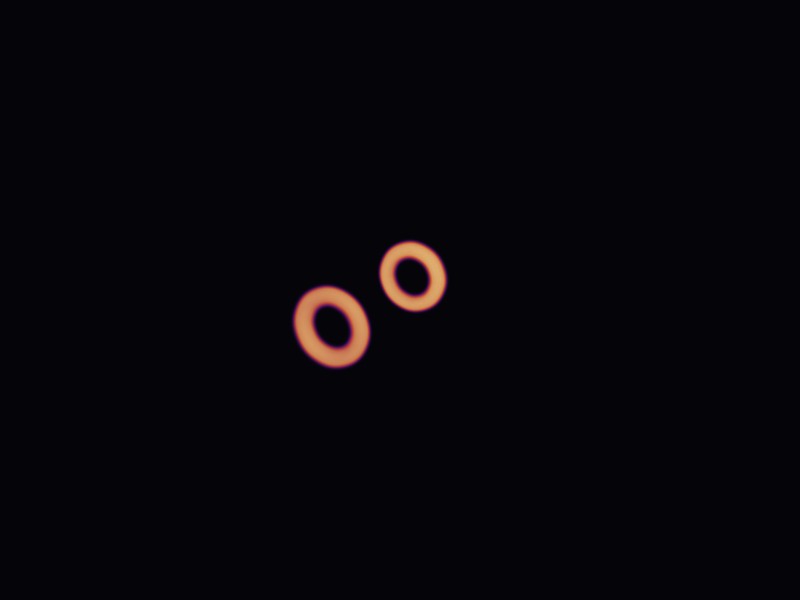}\hfill
\includegraphics[width=0.244\linewidth]{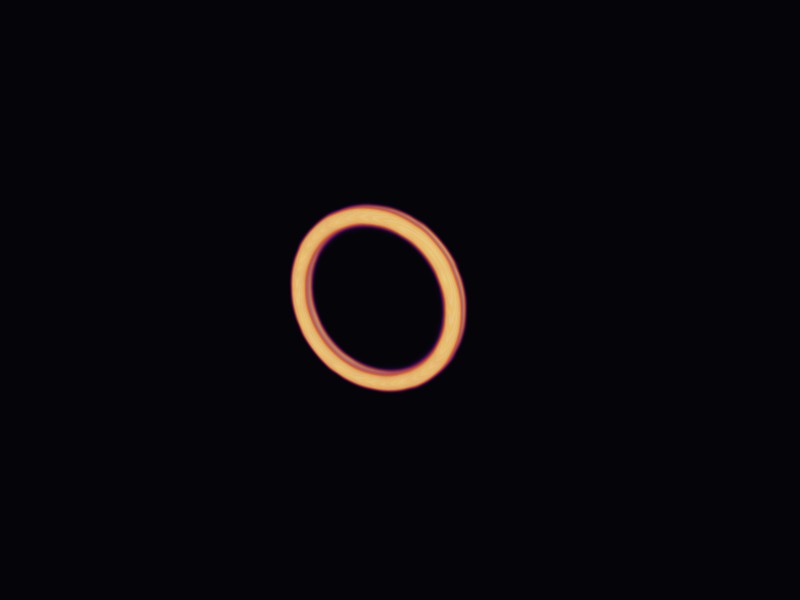}\hfill
\includegraphics[width=0.244\linewidth]{figures/simulations/vortex_ring_2.jpg}\hfill
\includegraphics[width=0.244\linewidth]{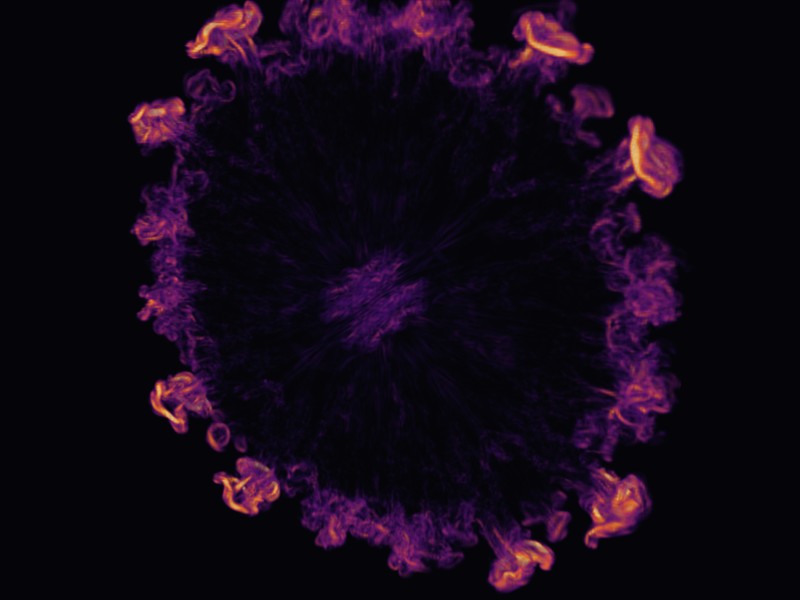}\\[-1pt]
\simlabels{$t=0.08$\,s}{$t=4.0$\,s}{$t=10.0$\,s}{$t=20.0$\,s}\\[5pt]
\textbf{(b)} Water drop (\nolinkurl{pic_flip})\\[2pt]
\includegraphics[width=0.244\linewidth]{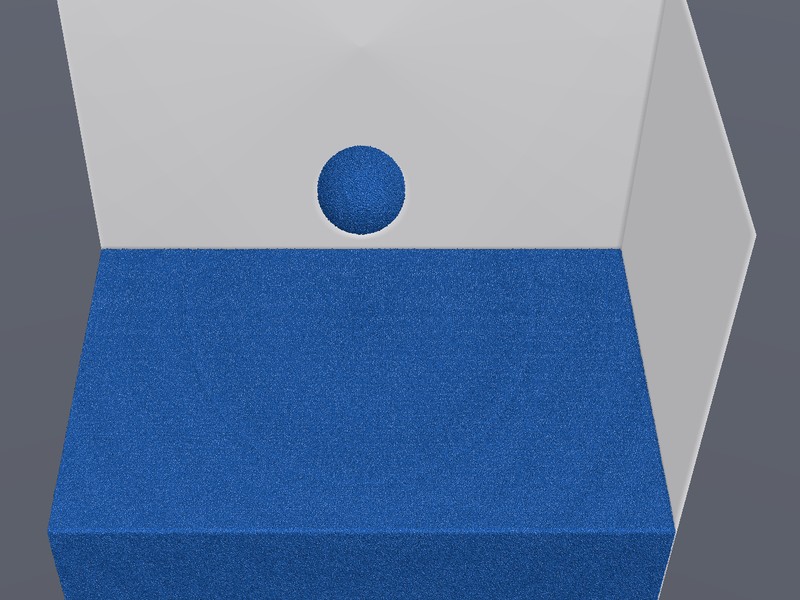}\hfill
\includegraphics[width=0.244\linewidth]{figures/simulations/water_drop_1.jpg}\hfill
\includegraphics[width=0.244\linewidth]{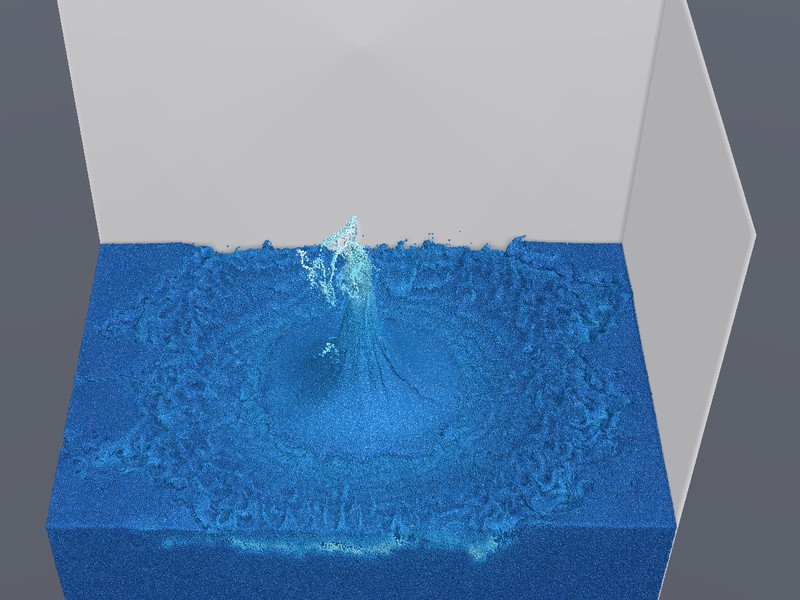}\hfill
\includegraphics[width=0.244\linewidth]{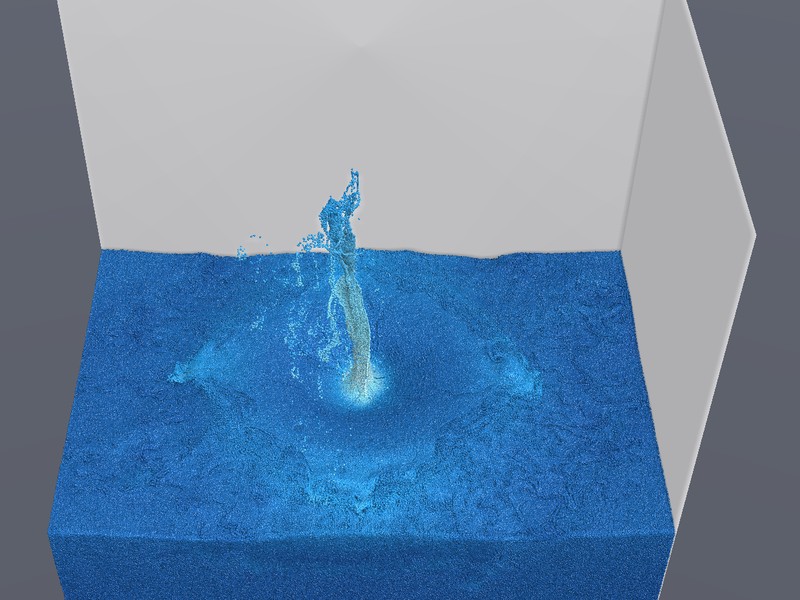}\\[-1pt]
\simlabels{$t=0$}{$t=0.50$\,s}{$t=0.83$\,s}{$t=1.17$\,s}\\[5pt]
\textbf{(c)} K\'arm\'an vortex street (\nolinkurl{stable_fluids})\\[2pt]
\includegraphics[width=0.244\linewidth]{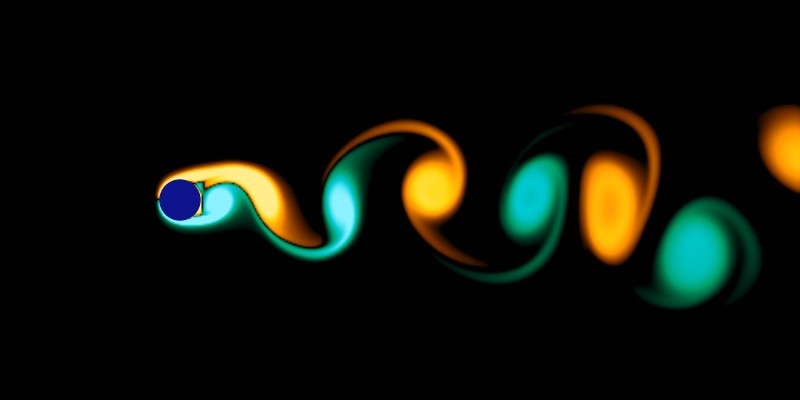}\hfill
\includegraphics[width=0.244\linewidth]{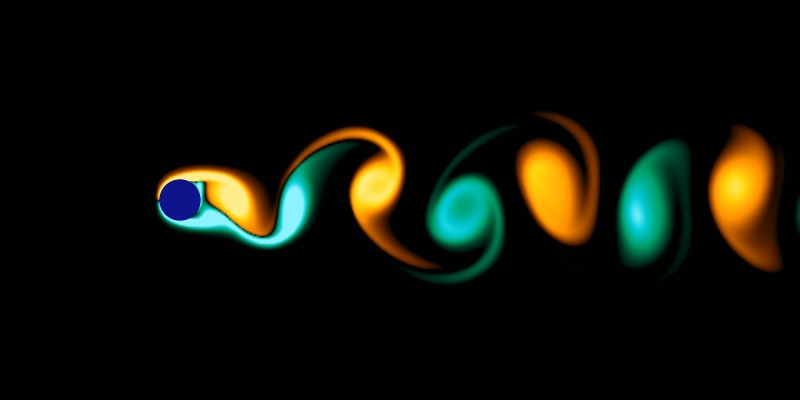}\hfill
\includegraphics[width=0.244\linewidth]{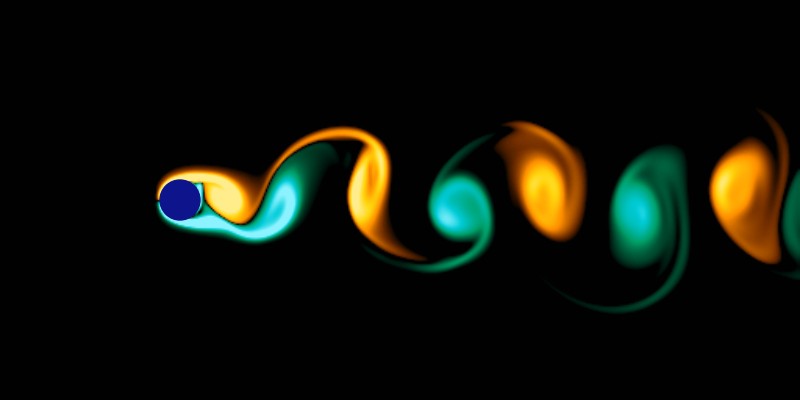}\hfill
\includegraphics[width=0.244\linewidth]{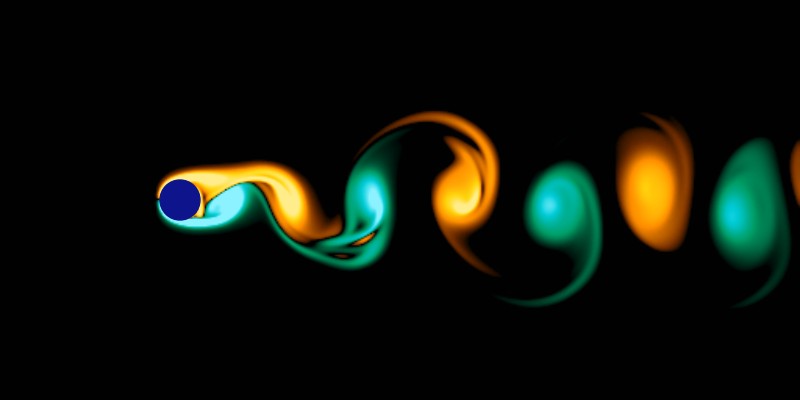}\\[-1pt]
\simlabels{$t=2.3$}{$t=4.7$}{$t=8.2$}{$t=12.0$}\\[5pt]
\textbf{(d)} Jelly cube (\nolinkurl{fem_explicit})\\[2pt]
\includegraphics[width=0.244\linewidth]{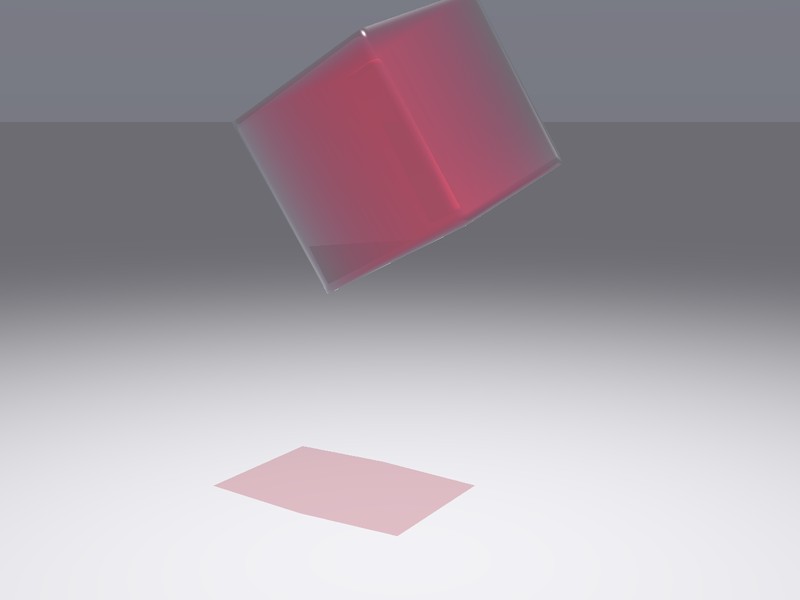}\hfill
\includegraphics[width=0.244\linewidth]{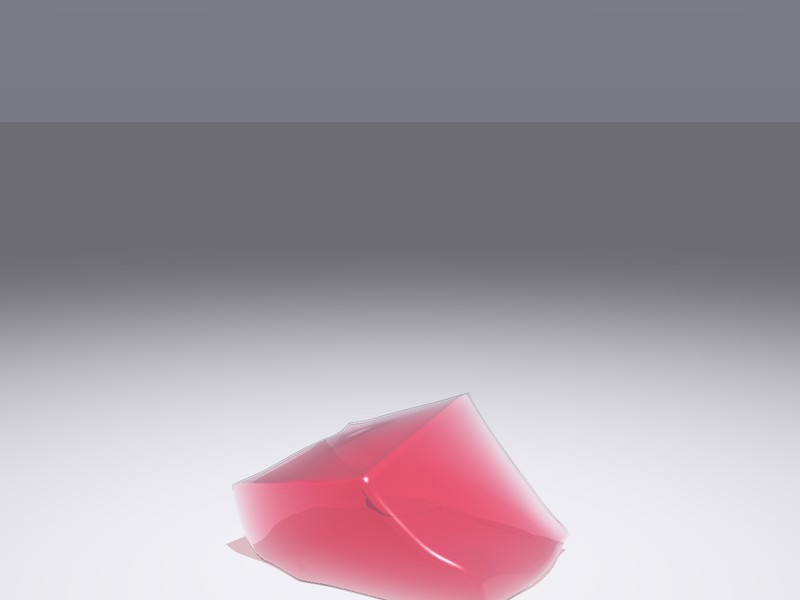}\hfill
\includegraphics[width=0.244\linewidth]{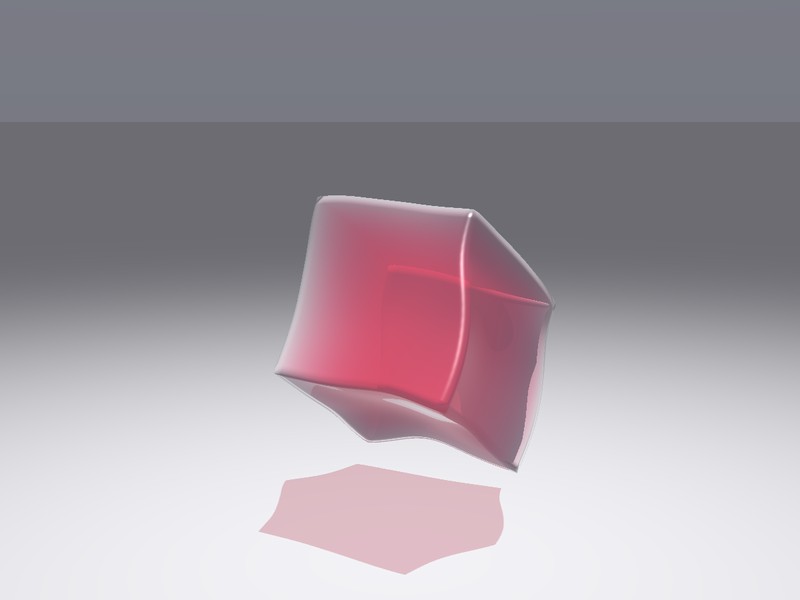}\hfill
\includegraphics[width=0.244\linewidth]{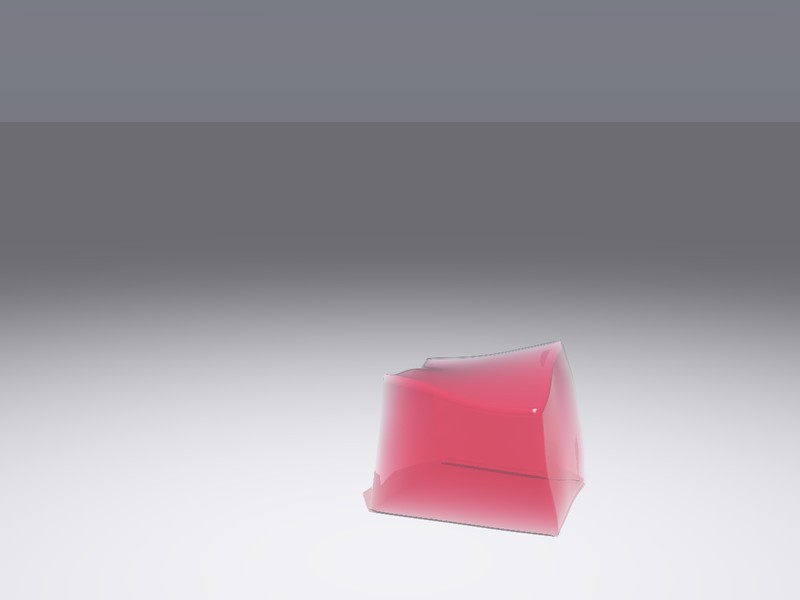}\\[-1pt]
\simlabels{$t=0$}{$t=0.42$\,s}{$t=0.75$\,s}{$t=2.0$\,s}\\[5pt]
\textbf{(e)} Snowball (\nolinkurl{mpm_explicit})\\[2pt]
\includegraphics[width=0.244\linewidth]{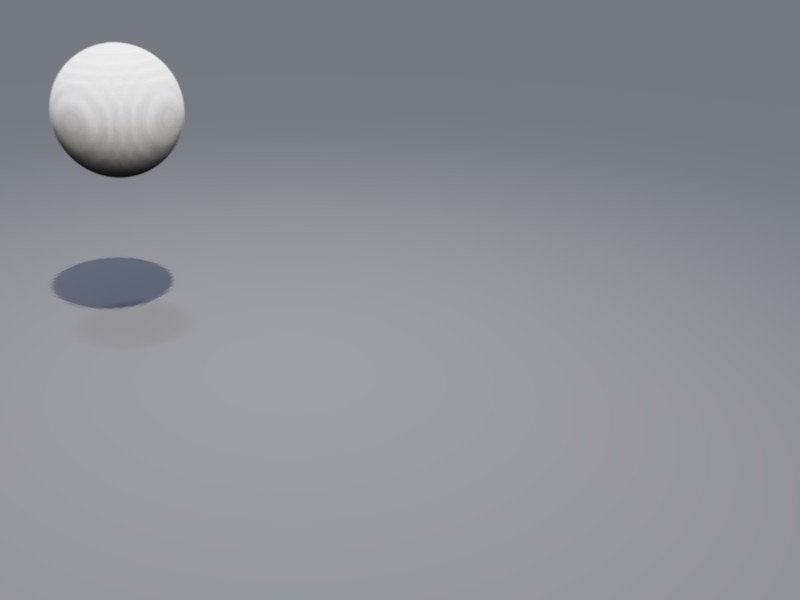}\hfill
\includegraphics[width=0.244\linewidth]{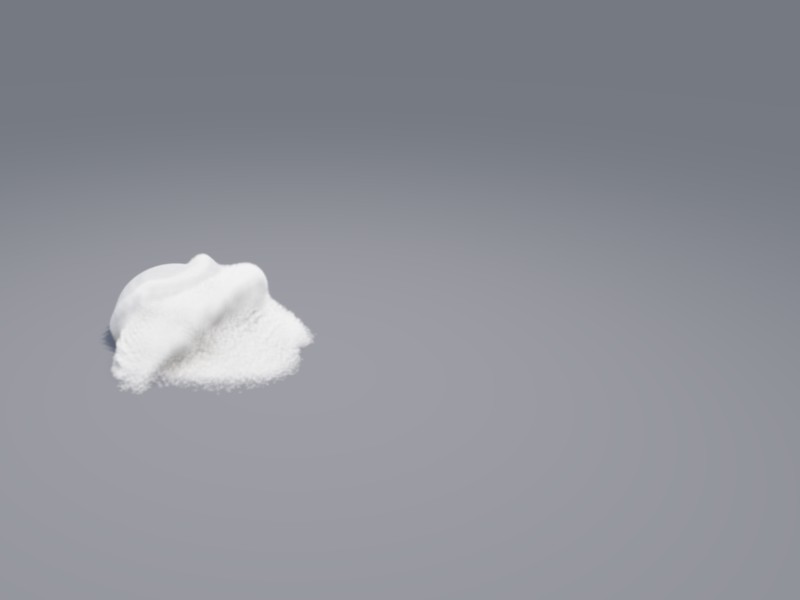}\hfill
\includegraphics[width=0.244\linewidth]{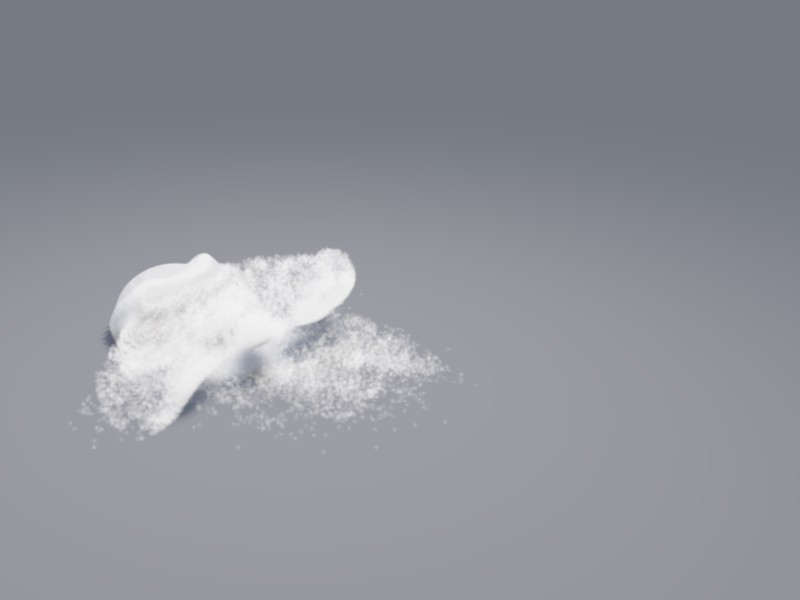}\hfill
\includegraphics[width=0.244\linewidth]{figures/simulations/snowball_3.jpg}\\[-1pt]
\simlabels{$t=0$}{$t=0.05$\,s}{$t=0.10$\,s}{$t=0.50$\,s}\\[5pt]
\textbf{(f)} Cloth on a sphere (\nolinkurl{xpbd})\\[2pt]
\includegraphics[width=0.244\linewidth]{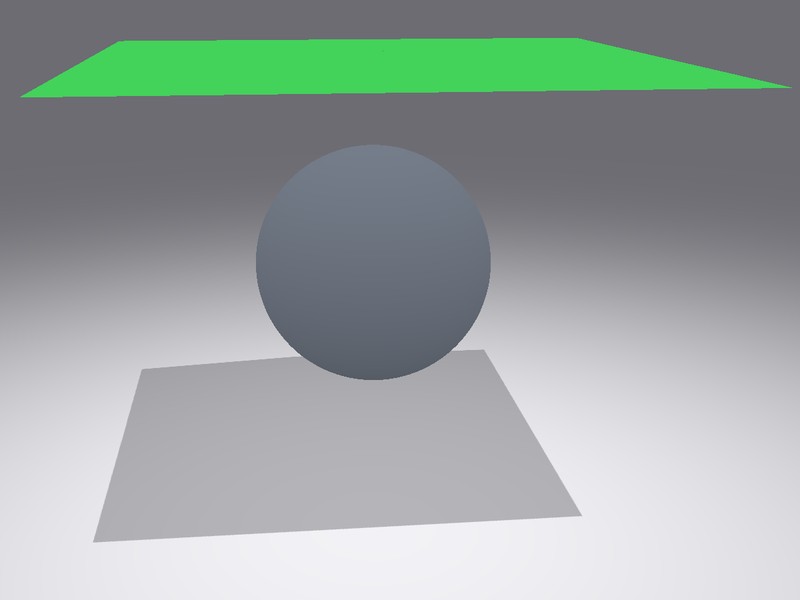}\hfill
\includegraphics[width=0.244\linewidth]{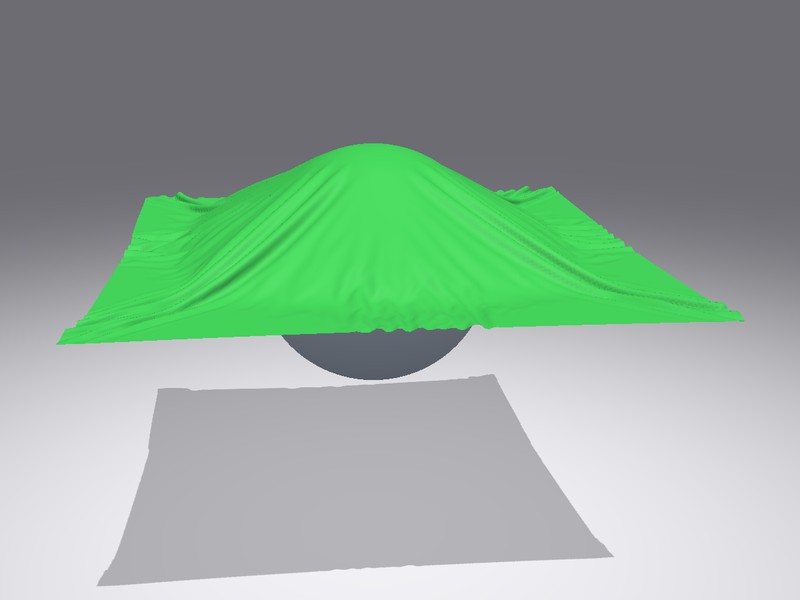}\hfill
\includegraphics[width=0.244\linewidth]{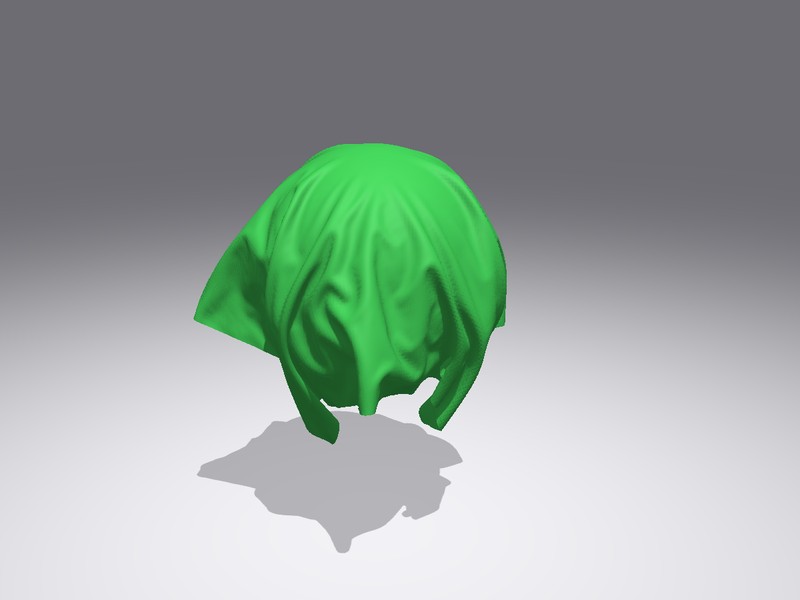}\hfill
\includegraphics[width=0.244\linewidth]{figures/simulations/cloth_3.jpg}\\[-1pt]
\simlabels{$t=0$}{$t=0.33$\,s}{$t=0.67$\,s}{$t=4.0$\,s}
\caption{Multi-step simulations driven by GPUPhysBench reference implementations, one per row, with the benchmark task named in parentheses. Each row shows four representative frames; times in (c) are in units of the channel height divided by the inflow speed.}
\label{fig:simulation_visualization}
\end{figure}

\end{document}